\documentclass[aps,twocolumn,floatfix]{revtex4}
\usepackage{epsfig}
\usepackage{graphicx}
\usepackage{dcolumn}
\usepackage{natbib}
\usepackage{amssymb}

\def\dj{\hbox{d\kern-0,347em \vrule width0,3em height1,252ex
depth-1,21ex \kern0,051em}}

\begin{document}

\title{Anomalous Roughness of Fracture Surfaces in 2D Fuse Models}

\author{Phani K.V.V. Nukala}
\affiliation{Computer Science and Mathematics Division,
Oak Ridge National Laboratory, Oak Ridge, TN 37831-6164, USA}
\author{Stefano Zapperi}
\affiliation{CNR-INFM, S3, Dipartimento di Fisica, Universit\`a di
Modena e Reggio Emilia, Via G. Campi 213A,  41100 Modena,
Italy}
\affiliation{ISI Foundation, Viale S. Severo 65, 10133 Torino,
Italy}
\author{Mikko J. Alava}
\affiliation{Laboratory of Physics, Helsinki University of Technology,
FIN-02015 HUT, Finland}
\author{Sr{\dj}an \v{S}imunovi\'{c}}
\affiliation{Computer Science and Mathematics Division,
Oak Ridge National Laboratory, Oak Ridge, TN~37831-6164, USA}

\begin{abstract}
We study anomalous scaling and multiscaling of
two-dimensional crack profiles in the random fuse model using both
periodic and open boundary conditions. Our large scale and
extensively sampled numerical results reveal the importance
of crack branching and coalescence of
microcracks, which induce jumps in the solid-on-solid crack
profiles. Removal of overhangs (jumps) in the crack profiles
eliminates the multiscaling observed in earlier
studies and reduces anomalous scaling. We 
find that the probability density distribution $p(\Delta
h(\ell))$ of the height differences $\Delta h(\ell) = [h(x+\ell) -
h(x)]$ of the crack profile obtained after removing the jumps in the
profiles has the scaling form $p(\Delta h(\ell)) = \langle\Delta
h^2(\ell)\rangle^{-1/2} ~f\left(\frac{\Delta h(\ell)}{\langle\Delta
h^2(\ell)\rangle^{1/2} }\right)$, and follows a Gaussian
distribution even for small bin sizes $\ell$. The anomalous scaling 
can be summarized with the scaling relation
$\left[\frac{\langle\Delta h^2(\ell)\rangle^{1/2}}{\langle\Delta
h^2(L/2)\rangle^{1/2}}\right]^{1/\zeta_{loc}} +
\frac{(\ell-L/2)^2}{(L/2)^2} = 1$, where $\langle\Delta
h^2(L/2)\rangle^{1/2} \sim L^{\zeta}$. 
\end{abstract}
%\PACS{62.20.Mk, 46.50.+a, 64.60.Ak}
\maketitle

\section{Introduction}
For over two decades, scaling of fracture surfaces has been a well
studied, yet a controversial issue \cite{breakdown,alava06}.
Experiments on several materials under different loading conditions
have shown that the fracture surface is self-affine \cite{man},
which implies that if the in plane length scales of a fracture
surface are scaled by a factor $\lambda$ then the out of plane
length scales (height) of the fracture surface scales by
$\lambda^\zeta$, where $\zeta$ is the roughness exponent. Many
experiments on several materials including metals \cite{metals},
glass \cite{glass}, rocks \cite{rocks} and ceramics \cite{cera} have
tested the scaling in three dimensions. The scaling regime in some
cases has been quite impressive, spanning five decades in metallic
alloys \cite{bouch}. It is an interesting question as to whether the
self-affinity measured so often can be replaced by more complicated
scenarios and whether in any particular setup and geometry the
exponents are universal as in the line-depinning scenario
\cite{ponson06ijf,ponson06}.

In two dimensions, recent studies have debated the picture of simple
self-affinity. In other words, for a two-dimensional crack profile $h(x)$ one can
look at various statistical measures including the dependence of the
roughness on sample/system size and the scaling of various moments
of $h(x)$. There is a discussion on whether the two-dimensional fracture
surfaces would exhibit self-affine or multi-affine scaling
\cite{procaccia,jstat2,bakke07,santucci07}. Ref. \cite{procaccia}
argues that a crack line $h(x)$ in two-dimensions is not
self-affine; instead, it exhibits a much complicated multi-affine
(or multiscaling) structure, with a non-constant scaling exponent
$\zeta_q$ for the $q$-th order correlation function $C_q(\ell) =
\langle |h(x+\ell)-h(x)|^q\rangle^{1/q} \sim \ell^{\zeta_q}$. In
analogy to kinetic roughening of surfaces, it has been argued that
fracture surfaces exhibit anomalous scaling \cite{anomalous}: the
{\it global} exponent describing the scaling of the crack width with
the sample size is larger than the local exponent measured on a
single sample \cite{exp-ano,exp-ano2}. This means that the typical
slope of $h(x)$ develops an algebraic dependence on the system size
$L$, and it is necessary to introduce two roughness exponents a
global one ($\zeta$) and a local one ($\zeta_{loc}$) whose
difference measures the $L$-dependent extra lengthscale.

In two dimensions, the available experimental results, mainly
obtained for paper samples, indicate a roughness exponent in the
range $\zeta \simeq 0.6-0.7$
\cite{jstat2,santucci07,kertesz93,engoy94,salminen03,rosti01}.
However, one should note that apparently one can measure for various
ordinary, industrial papers values that are significantly higher
than $\zeta=0.7$ \cite{menezessobrinho05}. The reasons for these
discrepancies are not clear. It has also been noted that the
roughness exponent is dependent on the crack velocity: at the onset
of fast crack propagation the exponent makes a small jump from its
value when the crack still grows in a stable fashion \cite{vanel}.

The theoretical understanding of the origin and universality of
crack surface roughness is often investigated by discrete lattice
(fuse, central-force, and beam) models. In these models the elastic
medium is described by a network of discrete elements such as fuses,
springs and beams with random failure thresholds. In the simplest
approximation of a scalar displacement, one recovers the random fuse
model (RFM) where a lattice of fuses with random threshold are
subject to an increasing external voltage \cite{deArcangelis85}.
Using two-dimensional RFM, the estimated crack surface roughness
exponents are: $\zeta = 0.7\pm0.07$ \cite{hansen91b},
$\zeta_{loc}=2/3$ \cite{sep-00}, and $\zeta = 0.74\pm0.02$
\cite{bakke}. Recently, using large system sizes (up to $L = 1024$)
with extensive sample averaging, we found that the crack roughness
exhibits anomalous scaling \cite{zns05}. The local and global
roughness exponents estimated using two different lattice topologies
are: $\zeta_{loc} = 0.72\pm0.02$ and $\zeta = 0.84\pm0.03$.
Anomalous scaling has been noted in the 3D numerical simulations as
well \cite{nukalapre3D}. The origins of anomalous scaling of
fracture surfaces is not yet clear although recent studies
\cite{jstat2,santucci07,bakke07} suggest that the origin of
multiscaling and anomalous scaling in numerical simulations may be
due to the existence of overhangs (jumps) in the crack profile.

In this paper, we further quantify the influence of these overhangs
in the crack profiles on multi-affine scaling and anomalous scaling
of crack roughness exponents. In particular, the questions we would
like to address in this article are the following: (i) whether
anomalous scaling of roughness observed in numerical simulations is
a result of these overhangs (or jumps) in the crack profiles, and
(ii) whether removing the jumps in the crack profiles completely
eliminates multiscaling. This should then imply a constant scaling
exponent $\zeta_{loc}$ such that the $q$-th order correlation
function $C_q(\ell) = \langle |h(x+\ell)-h(x)|^q\rangle^{1/q} \sim
\ell^{\zeta_{loc}}$. It should be noted that Gaussian distribution
for $p(\Delta h(\ell))$ has been noted in Refs.
\cite{salminen03,jstat2,santucci07,bakke07} only above a
characteristic scale where self-affine scaling of crack surfaces is
observed. In this study, we would like to further investigate
whether removing these jumps in the crack profiles extends the
validity of Gaussian probability density distribution $p(\Delta
h(\ell))$ of the height differences $\Delta h(\ell) = [h(x+\ell) -
h(x)]$ of the crack profile to even smaller window sizes $\ell$. We
also discuss the cases of open (OBC) and periodic boundary
conditions (PBC), since the presence of the former might have an
effect on whether "anomalous scaling" exists. For the PBC case, we show
that the crack profiles can be collapsed to a "semi-circle law", a scaling ansatz 
followed by many stochastic processes that return to the origin
\cite{baldassarri03}. The rest of the article consists of three
sections: first we introduce the numerical details. In Section III,
we go through all the numerical results, and finally Section IV
presents the conclusions.

\section{Model}
We consider numerical simulations using two-dimensional random fuse
model (RFM), where a lattice of fuses with random threshold are
subject to an increasing external voltage
\cite{deArcangelis85}. The lattice system we consider is a
triangular lattice of linear size $L$ with a central notch of length
$a_0$ (unnotched specimens imply $a_0 = 0$). All of the lattice
bonds have the same conductance, but the bond breaking thresholds,
$t$, are randomly distributed based on a thresholds probability
distribution, $p(t)$. The burning of a fuse occurs irreversibly,
whenever the electrical current in the fuse exceeds the breaking
threshold current value, $t$, of the fuse. Periodic boundary
conditions are imposed in the horizontal directions ($x$ direction)
to simulate an infinite system and a constant voltage difference,
$V$, is applied between the top and the bottom of the lattice system
bus bars.

A power-law thresholds distribution $p(t)$ is used
by assigning $t = X^D$, where $X \in [0,1]$ is a uniform
random variable with density $p_X(X) = 1$ and
$D$ represents a quantitative measure of disorder.
The larger $D$ is, the stronger the disorder. This results in
$t$ values between 0 and 1, with a cumulative
distribution $P(t) = t^{1/D}$. The average breaking
threshold is $<t> = 1/(D+1)$, and the probability that a fuse
will have breaking threshold less than the average breaking threshold $<t>$
is $P(<t>) = (1/(D+1))^{1/D}$. That is, the larger the $D$ is, the
smaller the average breaking threshold and the larger the probability
that a randomly chosen bond will have breaking threshold smaller than the
average breaking threshold.

Numerically, a unit voltage difference, $V = 1$, is set between the
bus bars (in the $y$ direction) and the Kirchhoff equations are solved to determine the
current flowing in each of the fuses. Subsequently, for each fuse $j$,
the ratio between the current $i_j$ and the breaking threshold $t_j$
is evaluated, and the bond $j_c$ having the largest value,
$\mbox{max}_j \frac{i_j}{t_j}$, is irreversibly removed (burnt).  The
current is redistributed instantaneously after a fuse is burnt
implying that the current relaxation in the lattice system is much
faster than the breaking of a fuse.  Each time a fuse is burnt, it is
necessary to re-calculate the current redistribution in the lattice to
determine the subsequent breaking of a bond.  The process of breaking
of a bond, one at a time, is repeated until the lattice system falls apart.

Using the algorithm proposed in Ref. \cite{nukalajpamg1}, we have performed
numerical simulation of fracture up to system
sizes $L = 512$ for unnotched samples and up to $L = 320$ for notched samples.
Our simulations cover an extensive parametric
space of ($L$, $D$ and $a_0$) given by: $L = \{64, 128, 192, 256,
320, 512\}$; $D = \{0.3, 0.4, 0.5, 0.6, 0.75, 1.0\}$; and
$a_0/L = \{0, 1/32, 1/16, 3/32, 1/8, 3/16, 1/4, 5/16, 3/8\}$. A minimum of 200
realizations have been performed for each case, but for many cases
2000 realizations have been used to reduce the statistical error.

\section{Crack Roughness}

\subsection{Crack width}
\begin{figure}[hbtp]
\includegraphics[width=8cm]{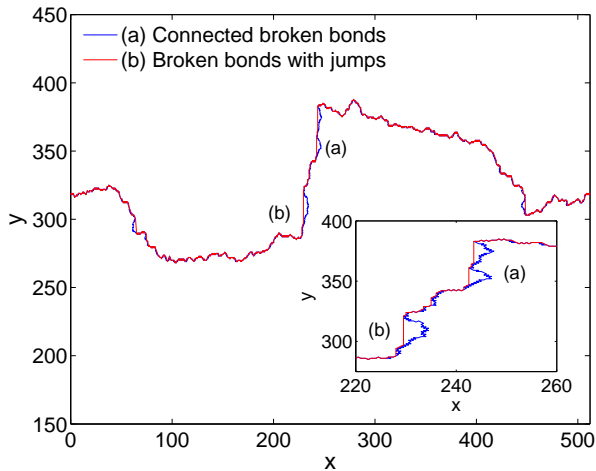}
\caption{(Color online) A typical crack in a fuse lattice system of
size $L \times L$ with $L = 512$. This crack, identified as (a) in
the figure has dangling ends, which are removed to obtain a single
valued crack profile $h(x)$, identified as (b) in the figure. This
final crack $h(x)$ possesses finite jumps that arise due to the
solid-on-solid projection to obtain a single-valued fracture
surface. The inset shows a zoomed portion of the crack.}
\label{fig:crack}
\end{figure}

Once the sample has failed, we identify the final crack, which
typically displays dangling ends and overhangs (see Fig.
\ref{fig:crack}). We remove them and obtain a single valued crack
line $h_x$, where the values of $x \in [0,L]$. For self-affine
cracks, the local width, $w(l)\equiv \langle \sum_x (h_x-
(1/l)\sum_X h_X)^2 \rangle^{1/2}$, where the sums are restricted to
regions of length $l$ and the average is over different
realizations, scales as $w(l) \sim l^\zeta$ for $l \ll L$ and
saturates to a value $W=w(L) \sim L^\zeta$ corresponding to the
global width. The power spectrum $S(k)\equiv \langle \hat{h}_k
\hat{h}_{-k} \rangle/L$, where $\hat{h}_k \equiv \sum_x h_x \exp
i(2\pi xk/L)$, decays as $S(k) \sim k^{-(2\zeta+1)}$. When anomalous
scaling is present \cite{anomalous,exp-ano,exp-ano2}, the exponent
describing the system size dependence of the surface differs from
the local exponent measured for a fixed system size $L$. In
particular, the local width scales as $w(\ell) \sim
\ell^{\zeta_{loc}}L^{\zeta-\zeta_{loc}}$, so that the global
roughness $W$ scales as $L^\zeta$ with $\zeta>\zeta_{loc}$.
Consequently, the power spectrum scales as $S(k) \sim
k^{-(2\zeta_{loc}+1)}L^{2(\zeta-\zeta_{loc})}$.

\begin{figure}[hbtp]
\includegraphics[width=8cm]{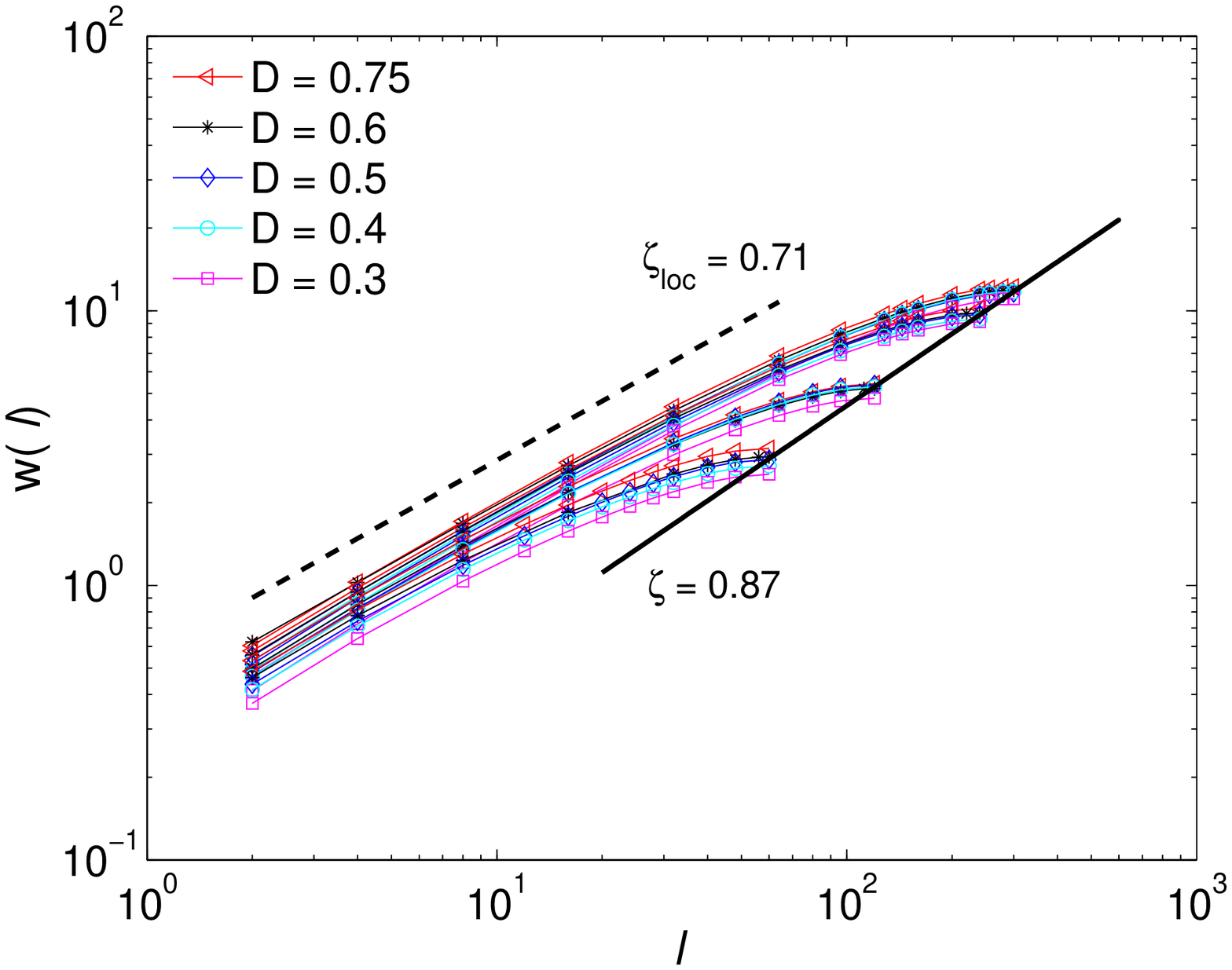}
\includegraphics[width=8cm]{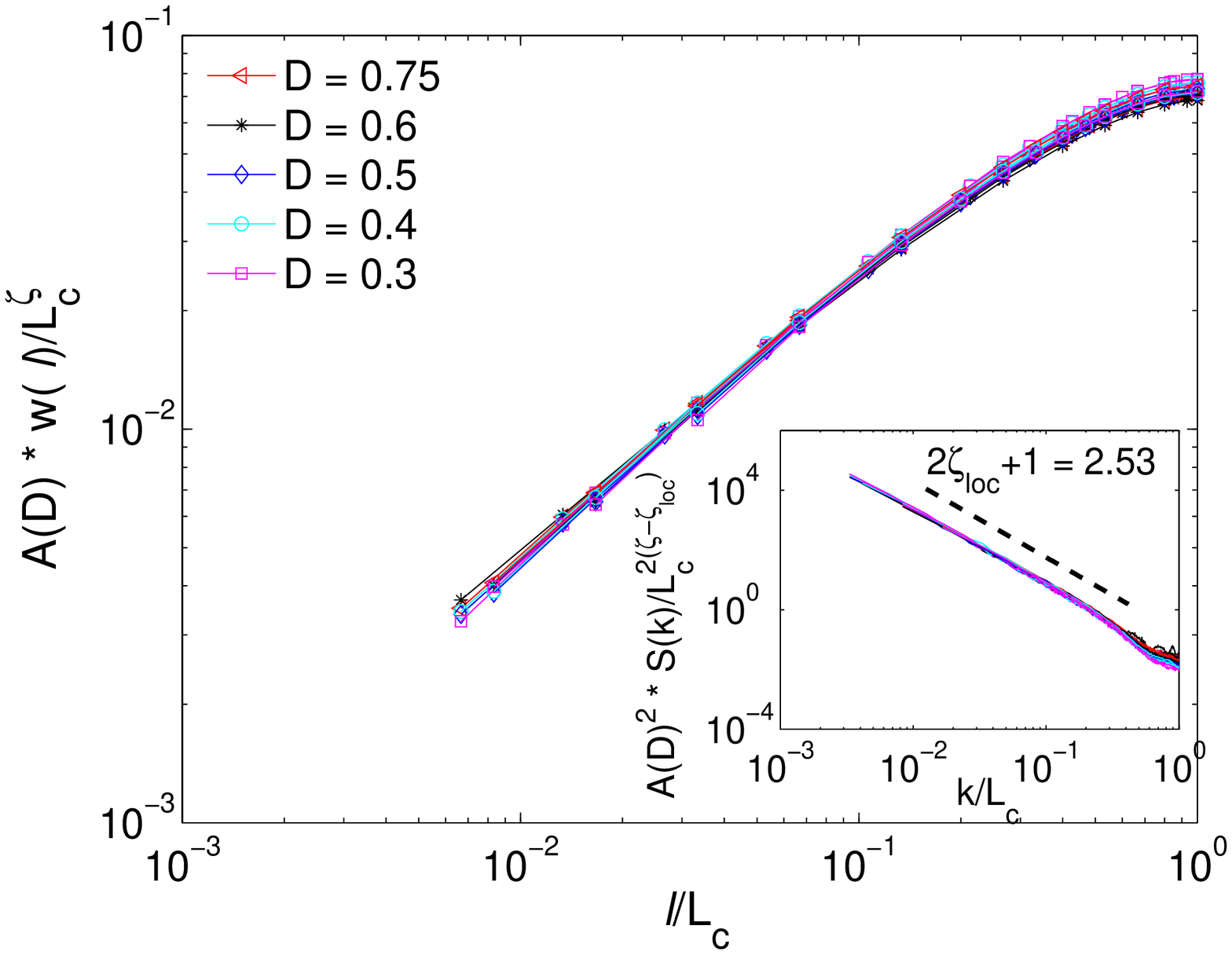}
\caption{(Color online) (a) Scaling of local and global widths
$w(l)$ and $W$ of the crack for different system sizes $L =
\{64,128,256,320\}$, disorder values $D$ and a fixed $a_0/L = 1/16$
value (top). The local crack width exponent $\zeta_{loc} = 0.71$ is
independent of disorder and differs considerably from the global
crack width exponent $\zeta = 0.87$. (b) Collapse of the crack width
data using the anomalous scaling law (bottom). $L_c = (L-a_0)$ is
the effective length of the crack profile. Collapse of the data
including a disorder dependent prefactor $A(D)$ for a given disorder
value implies that local and global roughness exponents are
independent of disorder. The inset shows collapse of power spectrum
$S(k)$ using the anomalous scaling law with $\zeta_{loc} = 0.71$ and
$\zeta = 0.87$. The slope in the inset defines the local exponent
via $-(2\zeta_{loc}+1) = -2.48$. (a)-(b) present a total of 20 data
sets.} \label{fig:widthD}
\end{figure}

Figure \ref{fig:widthD}a presents  the scaling of local and global
crack widths in systems with different disorder values and an
initial relative notch size of $a_0/L = 1/16$. The slopes of the
curves presented in Fig. \ref{fig:widthD}a suggest that a local
roughness exponent $\zeta_{loc} = 0.71$ that is independent of the
disorder. The global roughness exponent is estimated to be $\zeta =
0.87$, and differs considerably from the local roughness exponent
$\zeta_{loc}$. The collapse of the data in Fig. \ref{fig:widthD}b
clearly demonstrates that crack widths follow such an anomalous
scaling law. Notice that we have scaled away the amplitudes of the
roughness for all the different $D$ to achieve the maximal data
collapse to illustrate the universality. In the range of $D$
considered here the amplitudes vary by about 20 \%. The inset in
Fig. \ref{fig:widthD}b reports the data collapse of the power
spectra based on anomalous scaling for different disorder values.
This collapse of the data once again suggests that local roughness
is independent of disorder. A fit of the power law decay of the
spectrum yields a local roughness exponent of $\zeta_{loc}=0.74$.
This result is in close agreement with the real space estimate and
we can attribute the differences to the bias associated to the
methods employed \cite{sch-95}.

\subsection{Anomalous Scaling}
The scaling properties of the crack profiles $h(x)$ can also be
studied using the probability density distribution $p(\Delta
h(\ell))$ of the height differences $\Delta h(\ell) = [h(x+\ell) -
h(x)]$ of the crack profile between any two points on the reference
line ($x$-axis) separated by a distance $\ell$. Assuming the
self-affine property of the crack profiles implies that the
probability density distribution $p(\Delta h(\ell))$ follows the
relation
\begin{eqnarray}
p(\Delta h(\ell)) & \sim &  \langle\Delta h^2(\ell) \rangle^{-1/2}
~f\left(\frac{\Delta h(\ell)}{\langle\Delta h^2(\ell)\rangle^{1/2}
}\right) \label{pdelt}
\end{eqnarray}
where $\langle\Delta h^2(\ell)\rangle^{1/2}$ denotes  the width of
the height difference $\Delta h(\ell)$ over a length scale $\ell$.

Since for PBC the periodicity in crack profiles is analogous to
return-to-origin excursions arising in stochastic processes, we
propose the following ansatz for the local width $\langle\Delta
h^2(\ell)\rangle^{1/2}$ in height differences $\Delta h(\ell)$
\begin{eqnarray}
\langle\Delta h^2 (\ell)\rangle^{1/2} & = &  \langle\Delta
h^2(L/2)\rangle^{1/2} ~\phi\left(\frac{\ell}{L/2}\right)
\label{dhell}
\end{eqnarray}
with $\langle\Delta h^2(L/2)\rangle^{1/2} = L^\zeta$. The function
$\phi\left(\frac{\ell}{L/2}\right)$ is symmetric about $\ell = L/2$
and is constrained such that $\phi\left(\frac{\ell}{L/2}\right) = 0$
at $\ell = 0$ and $\ell = L$, and $\phi\left(\frac{\ell}{L/2}\right)
= 1$ at $\ell = L/2$. Based on these conditions, a scaling ansatz of
the form
\begin{eqnarray}
\left[\frac{\langle\Delta h^2(\ell)\rangle^{1/2}} {\langle\Delta
h^2(L/2)\rangle^{1/2}}\right]^{1/\zeta_{loc}} +
\frac{(\ell-L/2)^2}{(L/2)^2} & = & 1 \label{eq3}
\end{eqnarray}
similar to stochastic excursions or bridges can be proposed for
$\langle\Delta h^2(\ell)\rangle^{1/2}$, which implies a functional
form
\begin{eqnarray}
\phi\left (\frac{\ell}{L/2}\right) & = & \left[1 - \left(\frac{(\ell
- L/2)}{L/2}\right)^2\right]^{\zeta_{loc}} \label{dhell1}
\end{eqnarray}
for  $\phi\left(\frac{\ell}{L/2}\right)$ that is satisfied to a good
approximation by our numerical results. This scaling ansatz implies
anomalous scaling when $\zeta_{loc} \neq \zeta$. Upon further
simplication, Eq. (\ref{dhell1}) results in
\begin{eqnarray}
\phi\left(\frac{\ell}{L/2}\right)  & = & 4^{\zeta_{loc}}
\left(\frac{\ell}{L}\right)^{\zeta_{loc}} \left(1 -
\frac{\ell}{L}\right)^{\zeta_{loc}}  \label{dhell12}
\end{eqnarray}
which along with $\langle\Delta h^2(L/2)\rangle^{1/2}  = L^\zeta$
and Eq. (\ref{dhell}) illustrates how anomalous scaling appears in the scaling 
of local widths $\langle\Delta h^2(\ell)\rangle^{1/2}$, and how local
and global roughness exponents $\zeta_{loc}$ and $\zeta$ can be
computed based on numerical results.

\begin{figure*}[hbtp]
\begin{tabular}{cccc}
\includegraphics[width=6cm]{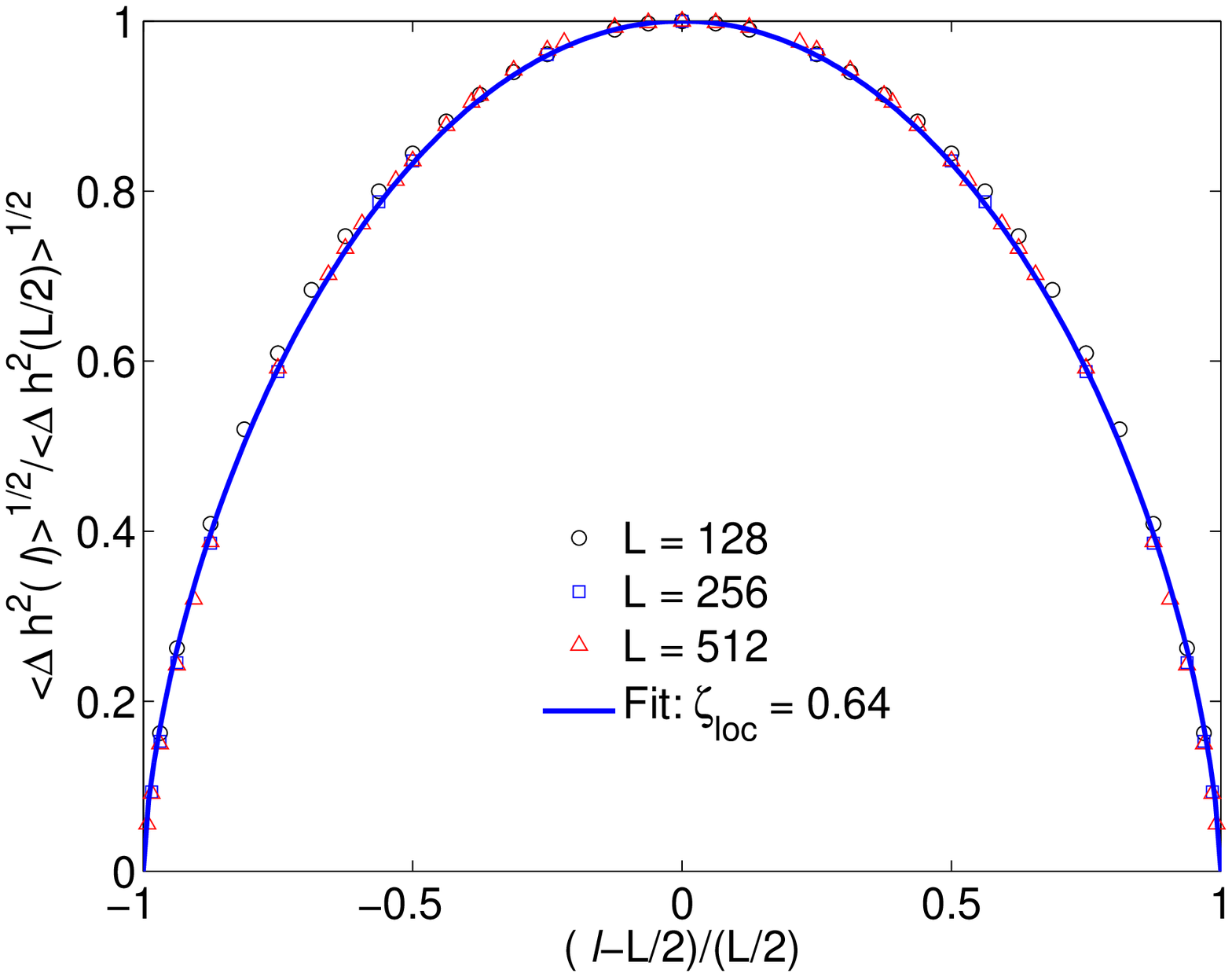} &
\includegraphics[width=6cm]{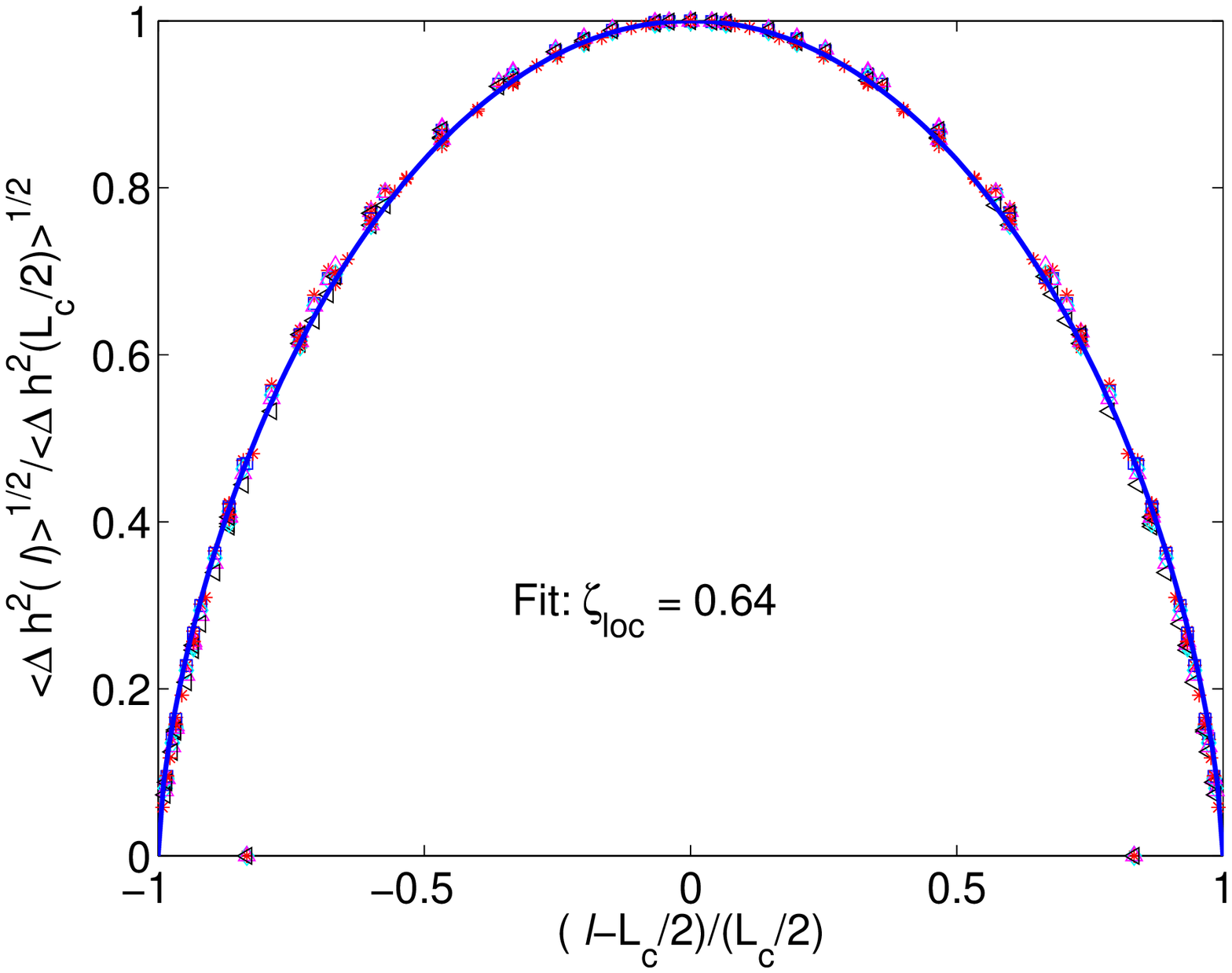} &
\includegraphics[width=6cm]{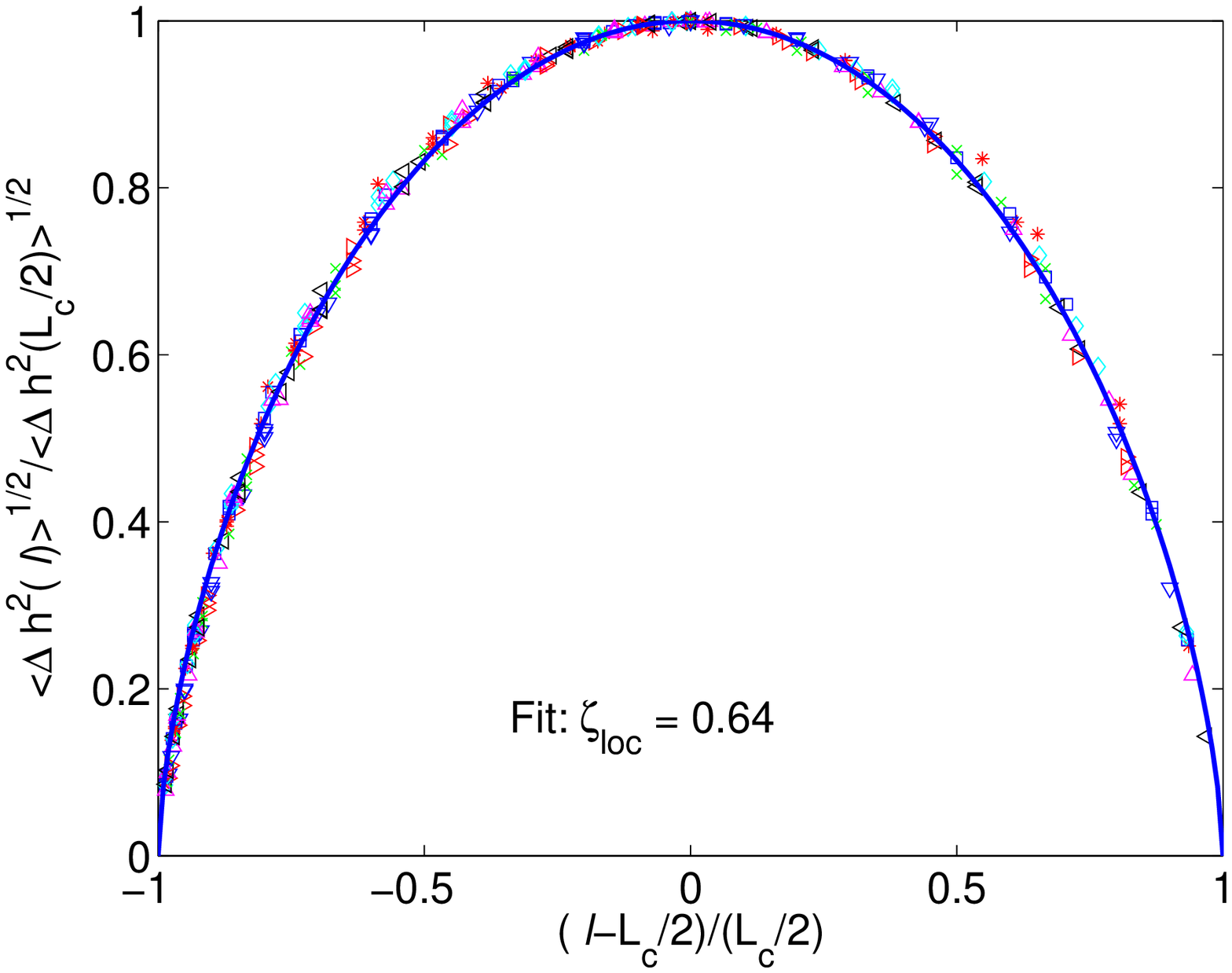}
\end{tabular}
\caption{(Color online) Scaling of $\langle\Delta
h^2(\ell)\rangle^{1/2}$ with window size $\ell$. (a) Un-notched fuse
simulations for $L = \{64,128,256,512\}$ and $D = 1.0$ (top left).
(b) Notched fuse simulations with various disorder values $D =
\{0.3,0.4,0.5,0.6,0.75\}$, system sizes $L =
\{64,128,192,256,320,512\}$, and a fixed notch size $a_0/L = 1/16$
(top middle). A total of eighteen data sets are plotted in figure.
(c) Notched fuse simulations for various relative crack sizes $a_0/L
= \{1/32,2/32,3/32,4/32,6/32,8/32,10/32,12/32\}$, system  sizes $L =
\{64,128,256,320\}$, and a fixed disorder of $D = 0.6$ (top right).
A total of twenty four data sets are plotted in figure.}
\label{fig:pbc_multi}
\end{figure*}

Figure \ref{fig:pbc_multi} presents the scaling of $\langle\Delta
h^2(\ell)\rangle^{1/2}$ based on the above ansatz (Eq. (\ref{eq3})).
The collapse of the $\langle\Delta
h^2(\ell)\rangle^{1/2}/\langle\Delta h^2(L/2)\rangle^{1/2}$ data for
different system sizes $L$ and window sizes $\ell$ onto a scaling
form given by Eq. (\ref{eq3})
 with $\zeta_{loc} = 0.64$ can be
clearly seen in Figs. \ref{fig:pbc_multi}(a)-(c). In particular,
Figs. \ref{fig:pbc_multi}(a)-(c) present the data for unnotched and
notched samples with varying amounts of disorder $D$ and relative
crack sizes $a_0/L$. The collapse of the data for varying amounts of
disorder ($0.3 \le D \le 1$) and relative crack sizes ($0 \le a_0/L
\le 3/8$) can be clearly seen in these figures and suggests that
local roughness exponent $\zeta_{loc}$ is independent of disorder,
at least for the disorder ranges considered here. It is an interesting question
as to why the $\zeta_{loc}$ from collapsing the average crack
profiles does not agree with the value from the local width,
demonstrated in Fig. \ref{fig:widthD}.

\begin{figure}[hbtp]
\includegraphics[width=8cm]{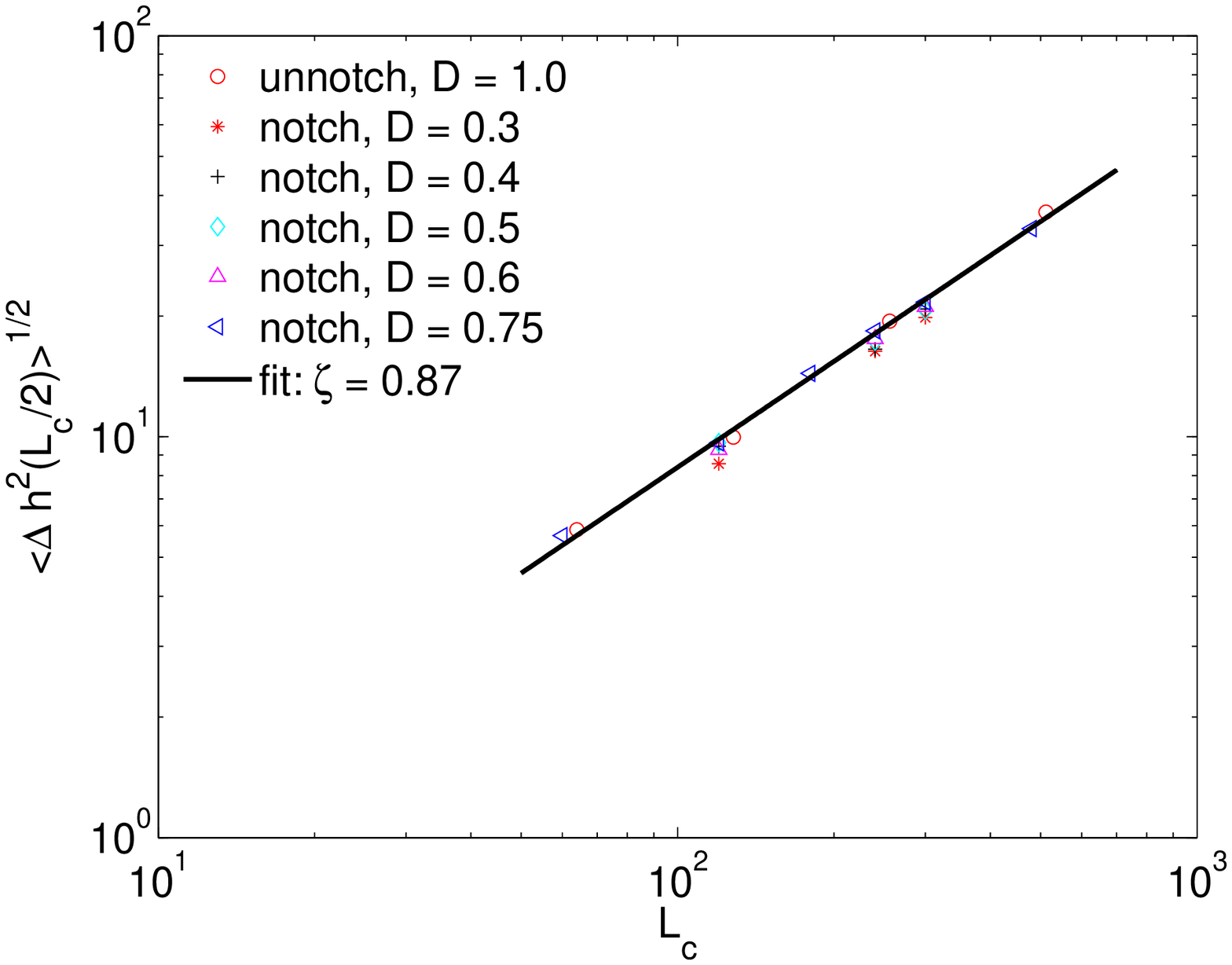}
\includegraphics[width=8cm]{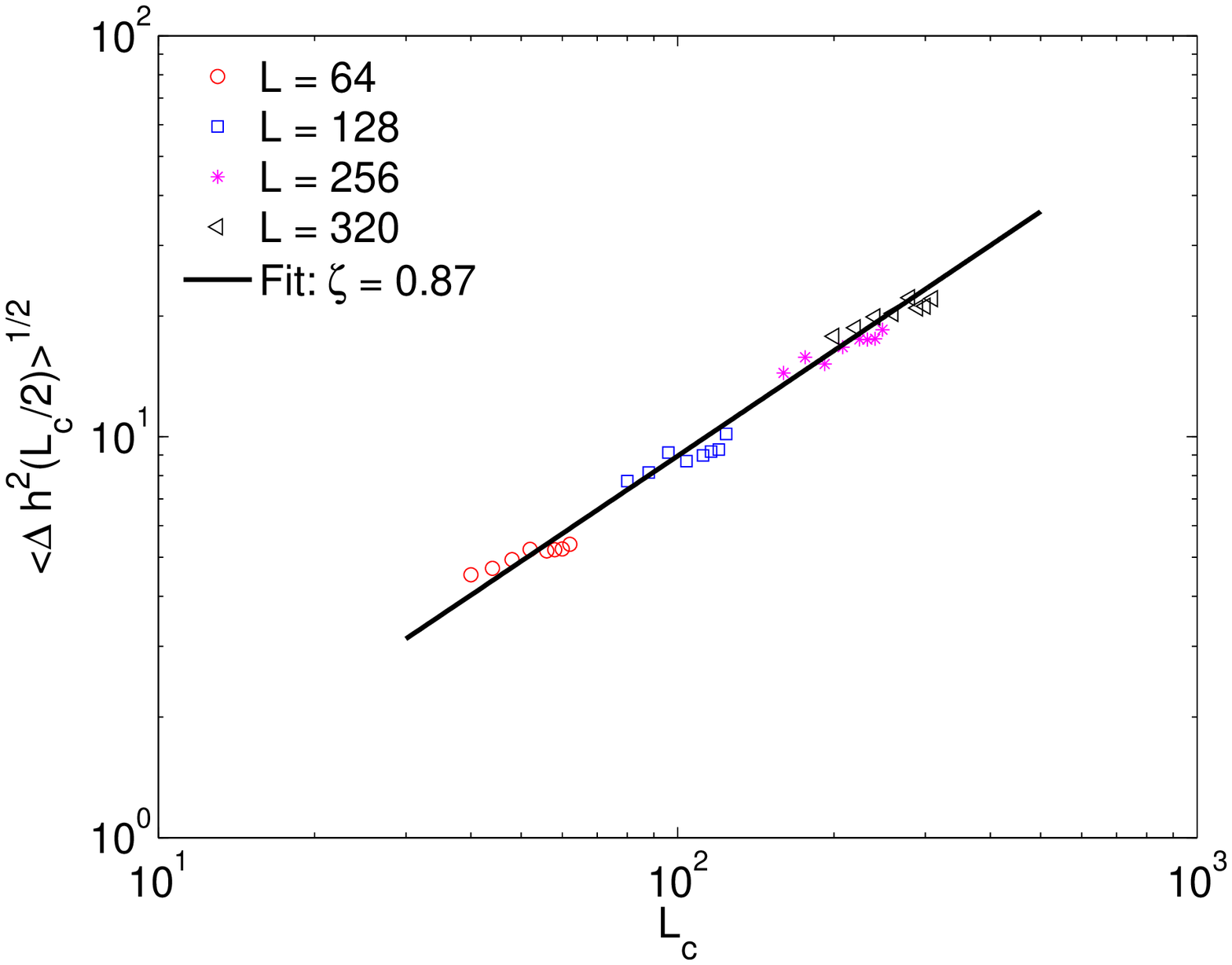}
\caption{(Color online) Scaling of $\langle\Delta
h^2(L/2)\rangle^{1/2}$ with system size $L$. For notched samples, we
use the effective length of the crack profile $L_c = L - a_0$. (a)
Scaling of $\langle\Delta h^2(L/2)\rangle^{1/2}$ is shown for
unnotched samples and for samples with a fixed relative notch size
of $a_0/L = 1/16$ having varying amounts of disorder $D$. (b)
Scaling of $\langle\Delta h^2(L_c/2)\rangle^{1/2}$ for samples with
varying notch sizes and a fixed disorder of $D = 0.6$ (bottom).}
\label{fig:width}
\end{figure}

Figures \ref{fig:width}(a)-(b) present the scaling of $\langle\Delta
h^2(L/2)\rangle^{1/2}$ for various notched and unnotched samples
with varying amounts of disorder and relative crack sizes. The data
presented in these figures shows that $\langle\Delta
h^2(L/2)\rangle^{1/2} \sim L^\zeta$ with $\zeta = 0.87$ in agreement
with the previously given value for the global width
exponent. In Figure \ref{fig:width}(b) one can note that
there is a $a_0/L$-dependent amplitude and the data follow the
0.87-exponent at fixed $a_0/L$. Since there exists a significant
difference between the global and local roughness exponents ($\zeta$
and $\zeta_{loc}$, respectively), here again we can conclude that
crack profiles obtained using the fuse models exhibit anomalous
roughness scaling.

\begin{figure}[hbtp]
\includegraphics[width=8cm]{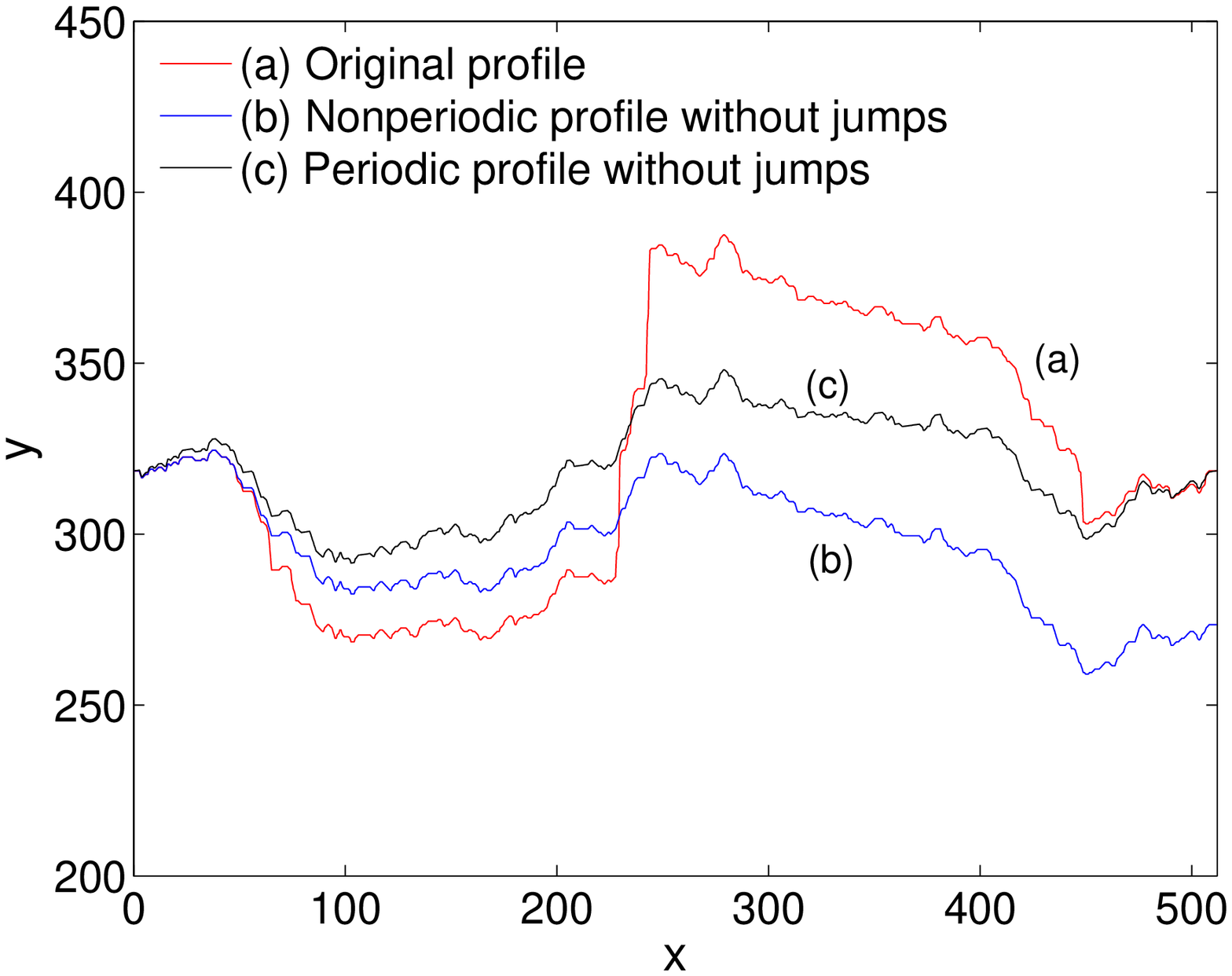}
\caption{(Color online) Figure shows a typical single valued crack
profile $h(x)$ with jumps based on solid-on-solid projection scheme
(identified as (a)). Removing the jumps in the crack profile $h(x)$
makes it a non-periodic profile (identified as (b)). Subtracting a
linear profile from this non-periodic  profile results in a periodic
profile (identified as (c)).} \label{fig:crack_nojumps}
\end{figure}

The questions that we would like to resolve in the following are
whether this anomalous scaling and multi-affine scaling of crack
surface roughness are a consequence  of the jumps in the crack
profiles induced by the crack overhangs (see Fig. \ref{fig:crack}).
As shown in Fig. \ref{fig:crack_nojumps}, removal of jumps from an
initially periodic crack profile $h(x)$ makes the resulting crack
profile $h_{NP}(x)$ nonperiodic, where the subscript $NP$ refers to
{\it nonperiodicity} of the profiles. A direct evaluation of the
roughness exponent using these nonperiodic profiles can be made.
However, such an evaluation of roughness exhibits finite size
effects for window sizes $\ell > L/2$. Alternatively, the roughness
of these resulting nonperiodic profiles can be evaluated by first
subtracting a linear profile $h_{lin}(x) = \left[h_{NP}(0) +
\frac{(h_{NP}(L)-h_{NP}(0))}{L} x\right]$ from the nonperiodic
profile $h_{NP}(x)$, and then evaluating the roughness of the
resulting periodic profile $h_{P}(x)$. In the following, we consider
the scaling of $h_{P}(x)$.

\begin{figure}[hbtp]
\includegraphics[width=8cm]{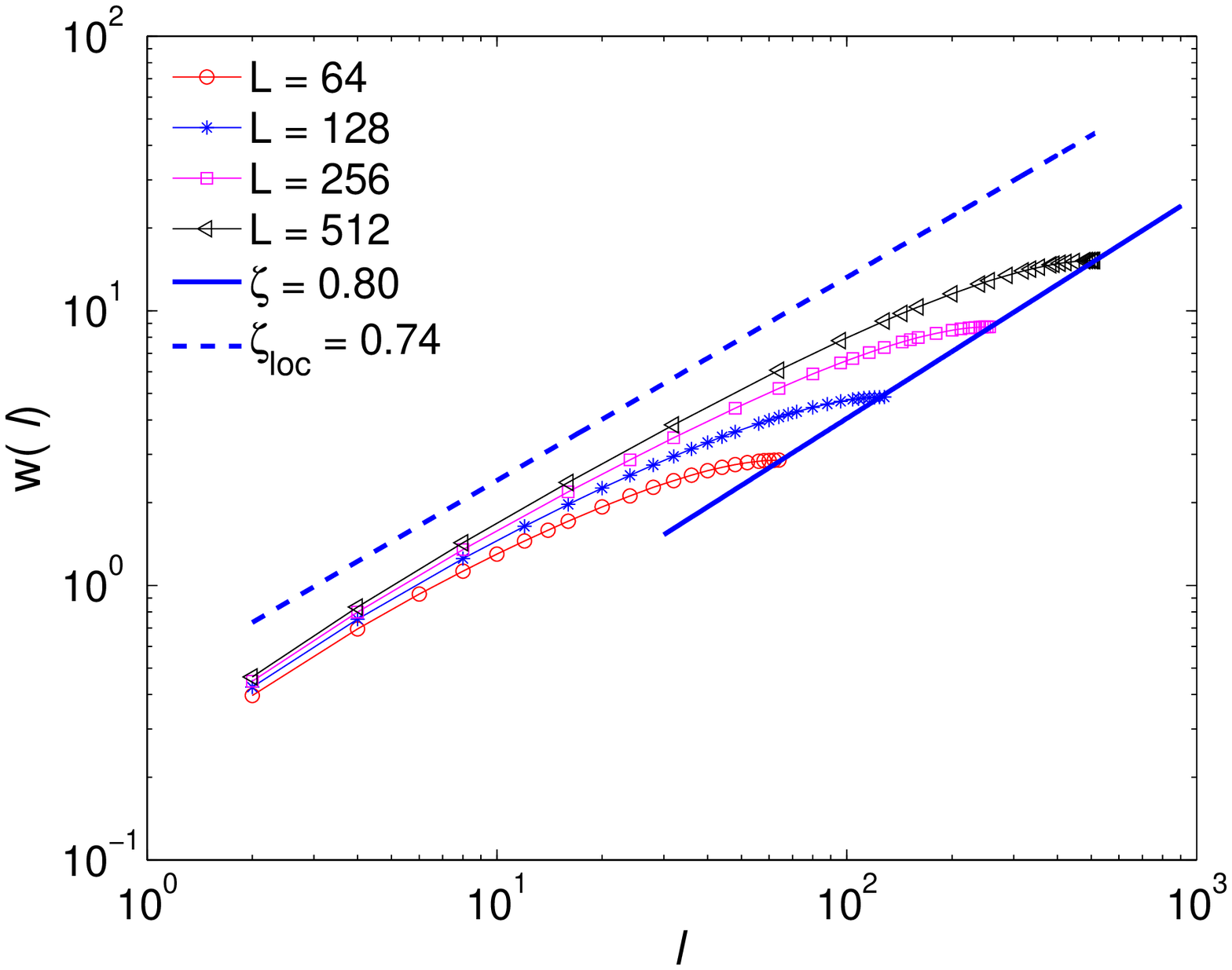}
\includegraphics[width=8cm]{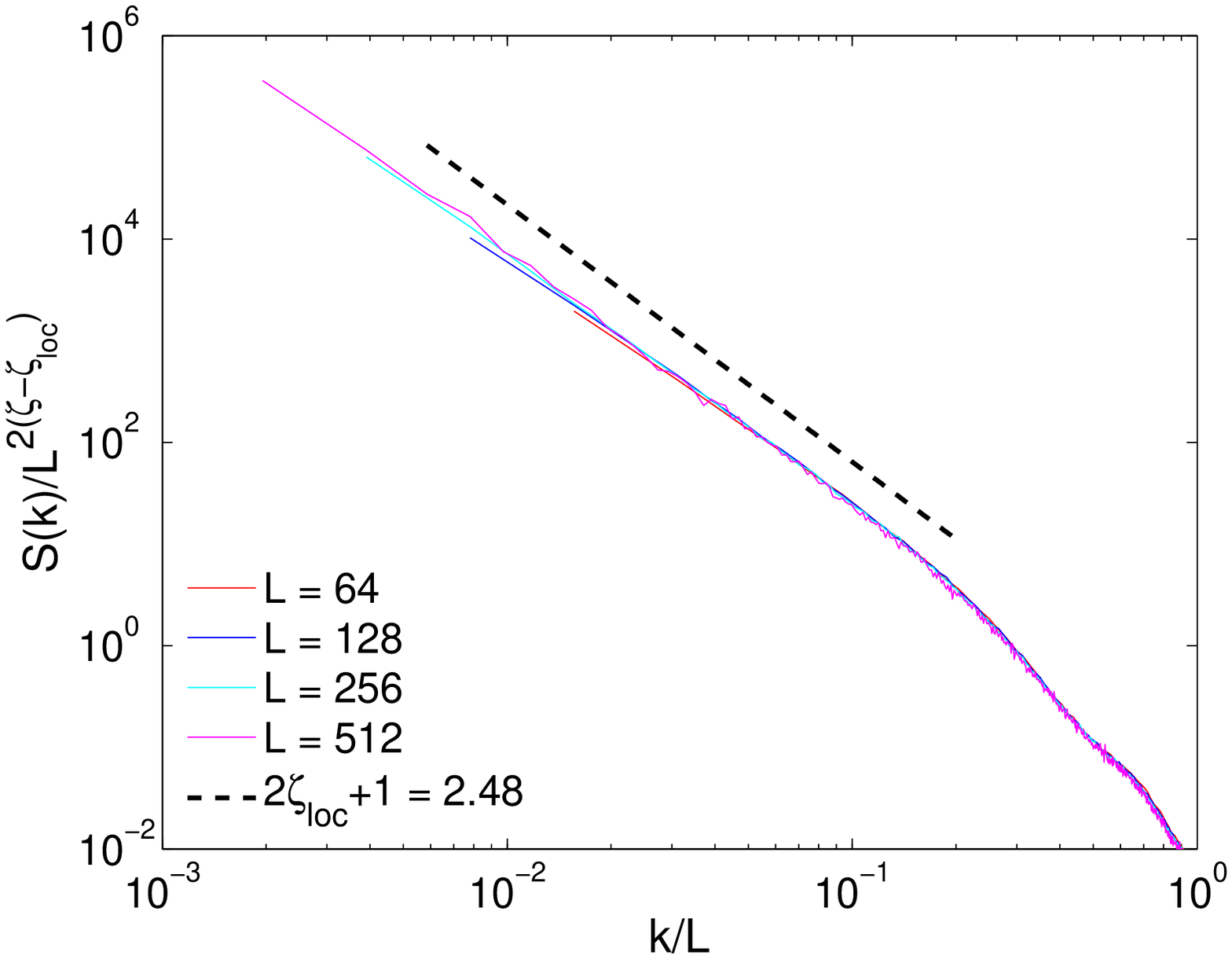}
\caption{(Color online) Scaling of crack profiles without the jumps.
(a) Scaling of crack width $w(\ell)$. The local and global
roughness exponents appear to be different although the difference
is considerably smaller than that with the jumps in the profiles.
(b) Scaling of power spectra of crack profiles. An excellent
collapse of the data is obtained using anomalous scaling law with
$2(\zeta-\zeta_{loc}) = 0.1$ and $\zeta_{loc} = 0.74$. Removal of
overhangs in the crack profiles does not appear to eliminate the
anomalous scaling of crack roughness in fuse models.}
\label{fig:pbc_width_nojumps}
\end{figure}

Figure \ref{fig:pbc_width_nojumps} presents the scaling of crack
width $w(\ell)$ with window size $\ell$ for crack profiles without
the jumps. The data presented in Fig. \ref{fig:pbc_width_nojumps}a
suggests that local and global roughness exponents ($\zeta_{loc} =
0.74$ and $\zeta = 0.80$) are not the same even after removing the
jumps in the crack profiles, although the difference between these
exponents is small. We have also investigated the power spectra
$S(k)$ of the crack profiles without the jumps in the crack profiles
(see Fig. \ref{fig:pbc_width_nojumps}b). An excellent collapse of
the data is obtained using the anomalous scaling law for power
spectrum with $2(\zeta-\zeta_{loc}) = 0.1$ and $\zeta_{loc} = 0.74$.
This result is consistent with the exponents measured using the
crack widths as in Fig. \ref{fig:pbc_width_nojumps}a.

\begin{figure}[hbtp]
%\begin{tabular}{cccc}
\includegraphics[width=8cm]{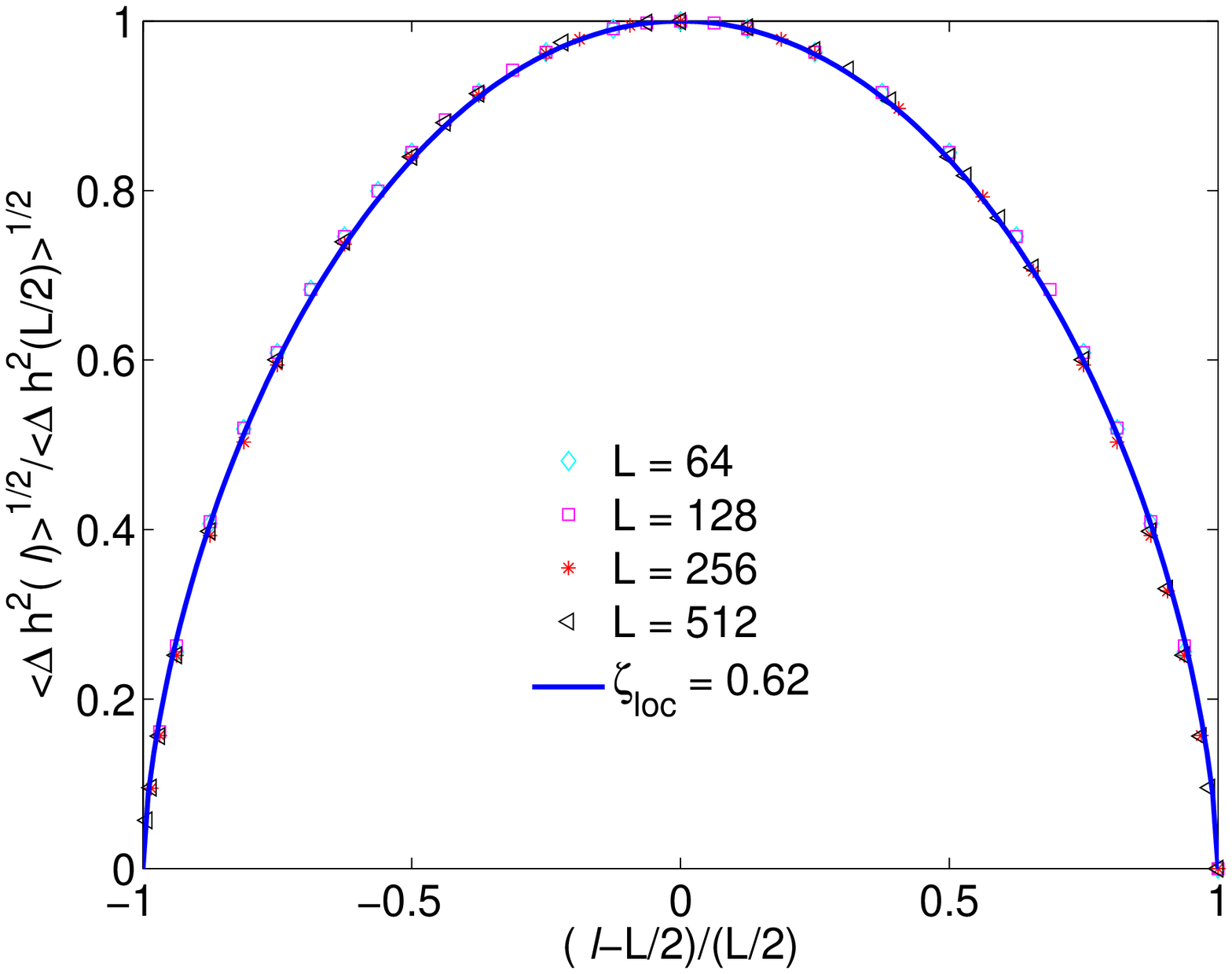} 
\includegraphics[width=8cm]{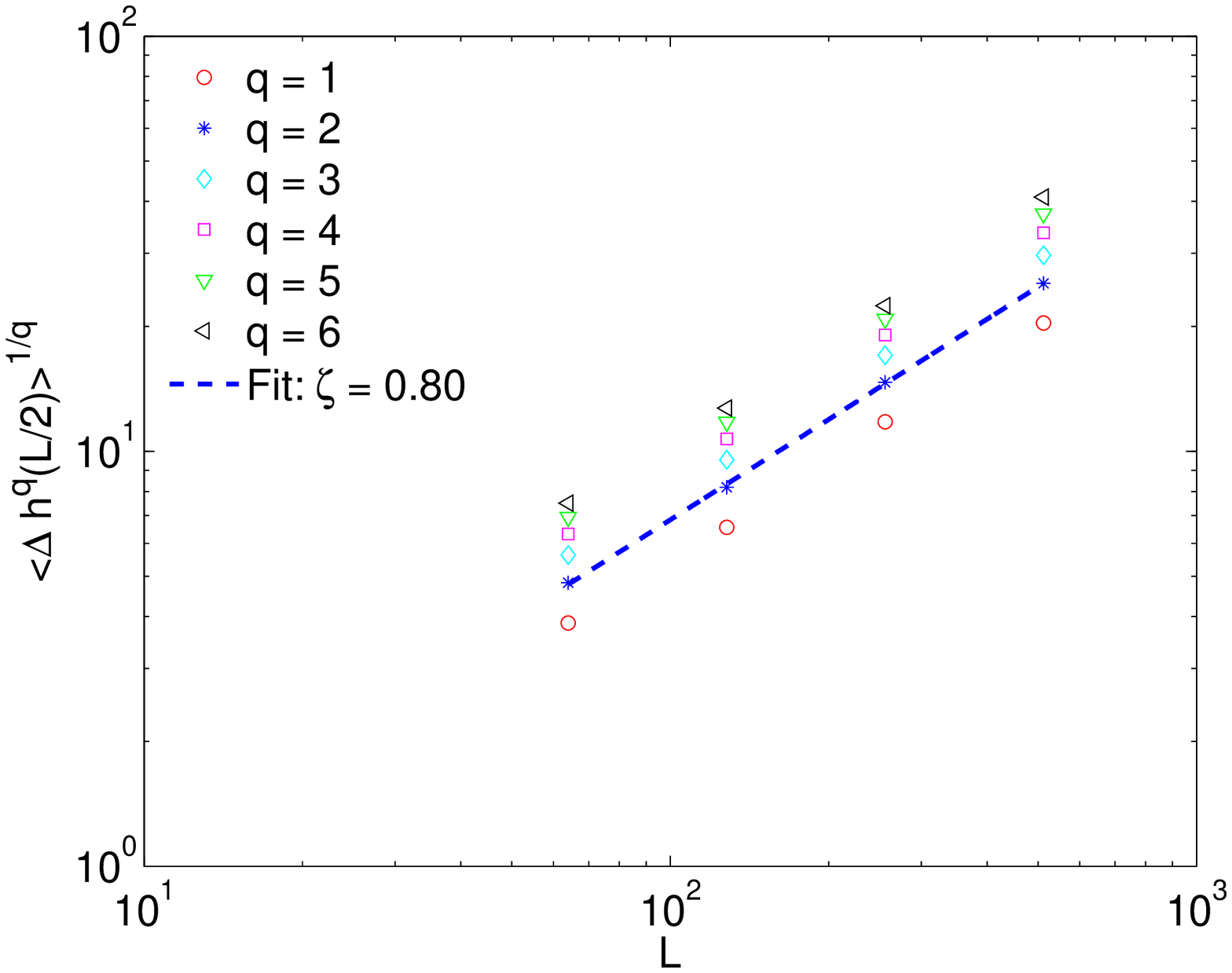}
\includegraphics[width=8cm]{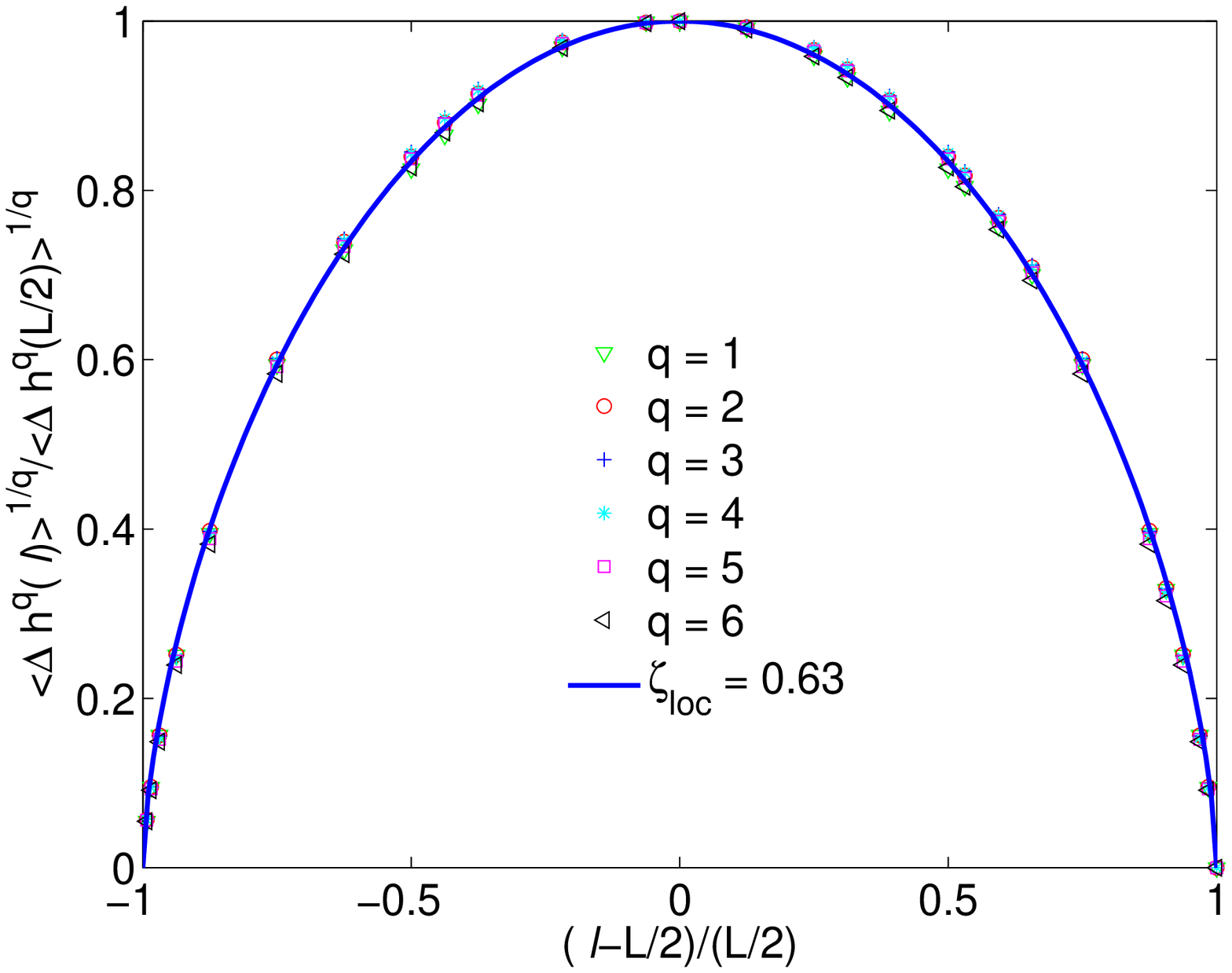} 
%\includegraphics[width=8cm]{height_multi_pbc2nopbc_nojumps_512.eps} \\
%\end{tabular}
\caption{(Color online) (a) Scaling of $\langle\Delta
h_{P}^2(\ell)\rangle^{1/2}$ with window size $\ell$ (top); 
(b) Scaling of $\langle\Delta h_P^2(L/2)\rangle^{1/2}$ with system size $L$ (middle); 
(c) Scaling of $\langle\Delta h_P^q(\ell)\rangle^{1/q}$ with window size $\ell$
for different moments $q$ of crack profiles without overhangs (bottom).}
\label{fig:pbc_multi_nojumps}
\end{figure}

In addition to the above two methods (crack width scaling and power
spectrum method) used for estimating the local and global roughness
exponents, we also used the scaling ansatz proposed in Eq.
(\ref{eq3}) to estimate the local and global roughness exponents.
Since the difference between the local and global exponents is
small, alternate ways of measuring these exponents provide a sense
of reliability into these estimates. Figures
\ref{fig:pbc_multi_nojumps}(a)-(c) present the scaling of crack
profiles $h_{P}(x)$. The collapse of the
$\langle\Delta h_P^2(\ell)\rangle^{1/2}/\langle\Delta
h_P^2(L/2)\rangle^{1/2}$ data for different system sizes $L$ and
window sizes $\ell$ onto a scaling form given by Eq. (\ref{eq3})
with $\zeta_{loc} = 0.62$ can be clearly seen in Fig.
\ref{fig:pbc_multi_nojumps}(a). This value is fairly close to the
0.64 quoted before. In addition, the collapse of the data presented
in Fig. \ref{fig:pbc_multi_nojumps}(c) for $\langle\Delta
h_P^q(\ell)\rangle^{1/q}/\langle\Delta h_P^q(L/2)\rangle^{1/q}$
demonstrates that multi-affine scaling of fracture surfaces arises
because of overhangs (jumps) with certain statistics in the crack
profile and removal of these jumps in the crack profiles completely
eliminates multiscaling of fracture surfaces.

The simple scaling (no  multiaffinity) is also evident through the
scaling of $\langle\Delta h_P^q(L/2)\rangle^{1/q}$ presented in
Fig. \ref{fig:pbc_multi_nojumps}(b). The slopes of the data for
moments $q = 1$ to $6$ of $\Delta h_P(L/2)$ are identical. An
interesting observation to be made is that $\langle\Delta
h_P^q(L/2)\rangle^{1/q} \sim L^\zeta$ with $\zeta = 0.80$ whereas
the local roughness exponent as obtained from Figs.
\ref{fig:pbc_multi_nojumps}(a) and (c) is $\zeta_{loc} = 0.62$. A
similar behavior is observed even when the linearity in the profile
is not subtracted: the scaling of these nonperiodic profiles
$h_{NP}(x)$ is in agreement with that obtained for periodic profiles
for window size $\ell \le L/2$ although finite size effects are
observed when window sizes $\ell > L/2$ are considered. The difference in these
exponents even after removing the jumps caused by overhangs in the
crack profile indicates that anomalous scaling is present in
two-dimensional fracture simulations using the fuse models and this
anomalous scaling is not due to the jumps in the crack profiles.

\subsection{The case of open boundaries}
\begin{figure}[hbtp]
\includegraphics[width=8cm]{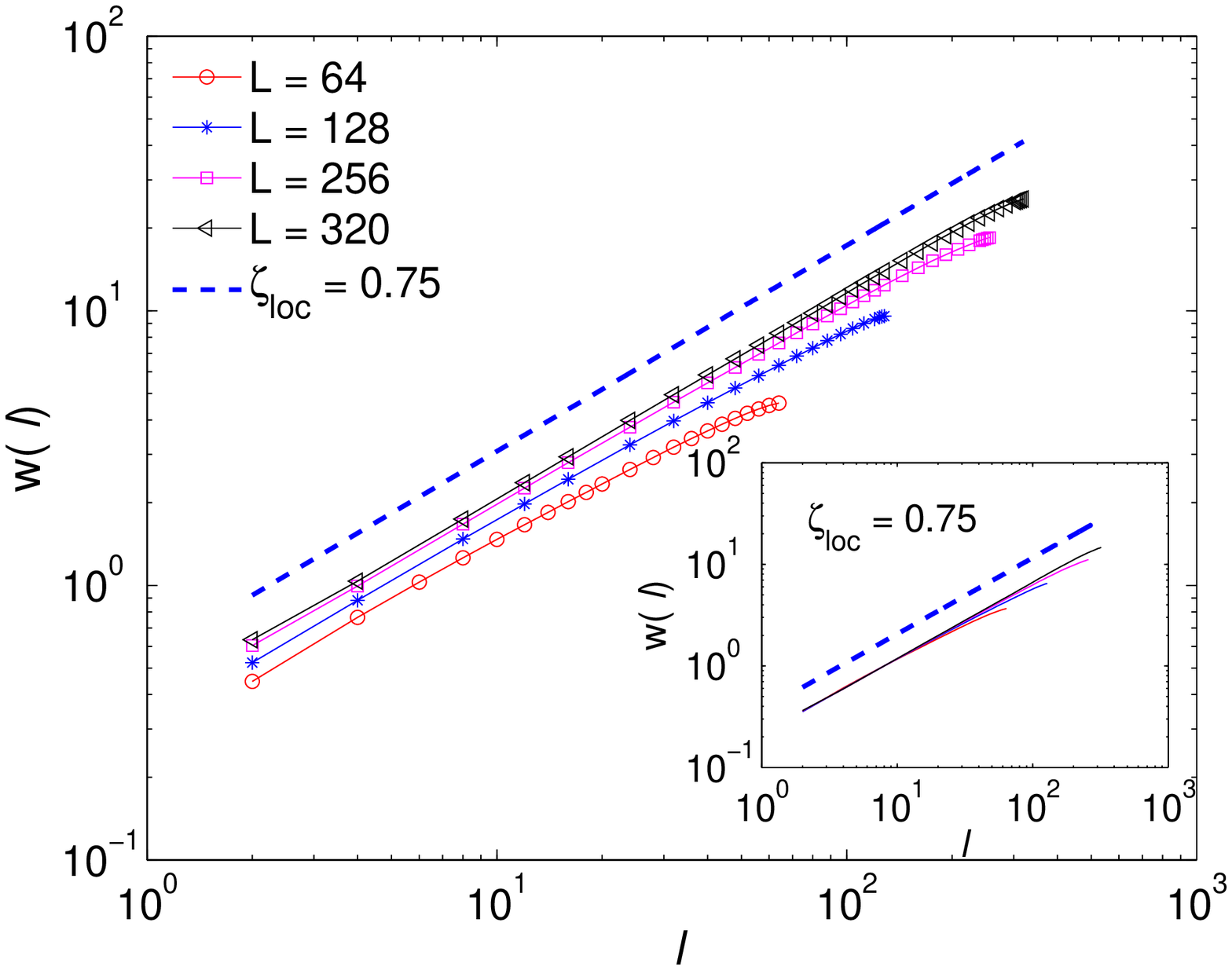}
\includegraphics[width=8cm]{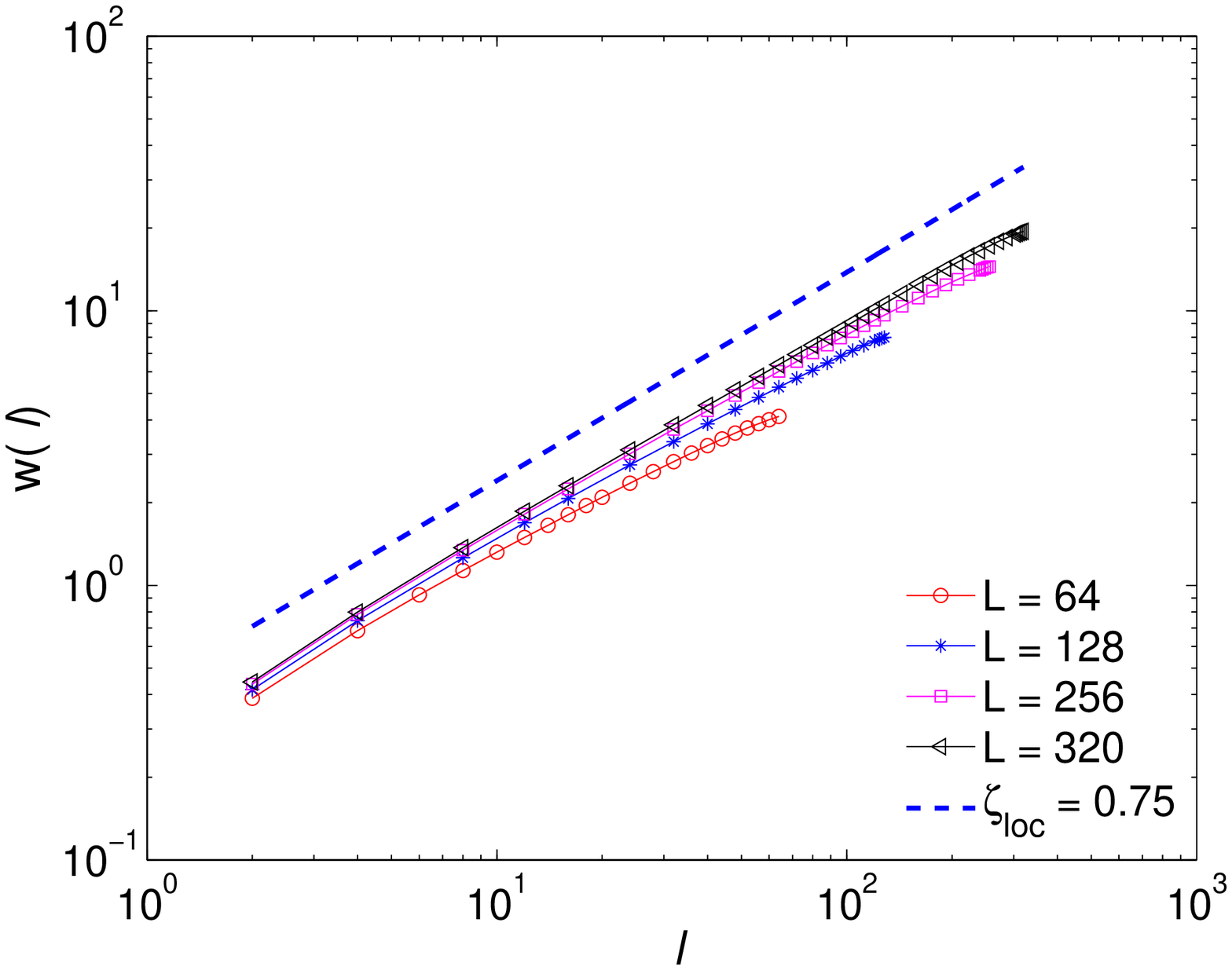}
\caption{(Color online) Scaling of crack width $w(\ell)$ with window
size $\ell$ for open boundary conditions. (a) Scaling of $w(\ell)$
for crack profiles with jumps. The inset presents the data shown in
the main figure after a $L$-dependent shift is applied. A power law
fit to the data estimates the local roughness exponent to be
$\zeta_{loc} = 0.75$. (b) Scaling of $w(\ell)$ for crack profiles
obtained after removing the jumps. plots in figure (b) indicate that
even with the removal of overhangs in the crack profile does not
eliminate this apparent anomalous scaling of crack roughness.}
\label{fig:obc_width}
\end{figure}

It is interesting to compare the PBC case with that of open boundary
conditions. Figure \ref{fig:obc_width} presents the scaling of crack
widths for fuse lattice simulations with open boundary conditions.
The data in Fig. \ref{fig:obc_width}a indicates that local roughness
exponent is $\zeta_{loc} = 0.75$. However, the data for different
system sizes does not collapse, which is an indication of anomalous
scaling. The inset in Fig. \ref{fig:obc_width}a shows that a simple
L-dependent shifting of the data achieves a perfect collapse of the data with
possible finite size deviations for window sizes $\ell$ approaching
the system size. 
Figure \ref{fig:obc_width}b presents the scaling of
$w(\ell)$ for crack profiles without the jumps. Even after removing
the jumps from the crack profiles, the crack widths data does not
collapse onto a single curve. This suggests that removal of
overhangs in the crack profile does not eliminate this apparent
anomalous scaling of crack roughness even for open boundary
conditions.

\begin{figure}[hbtp]
\includegraphics[width=8cm]{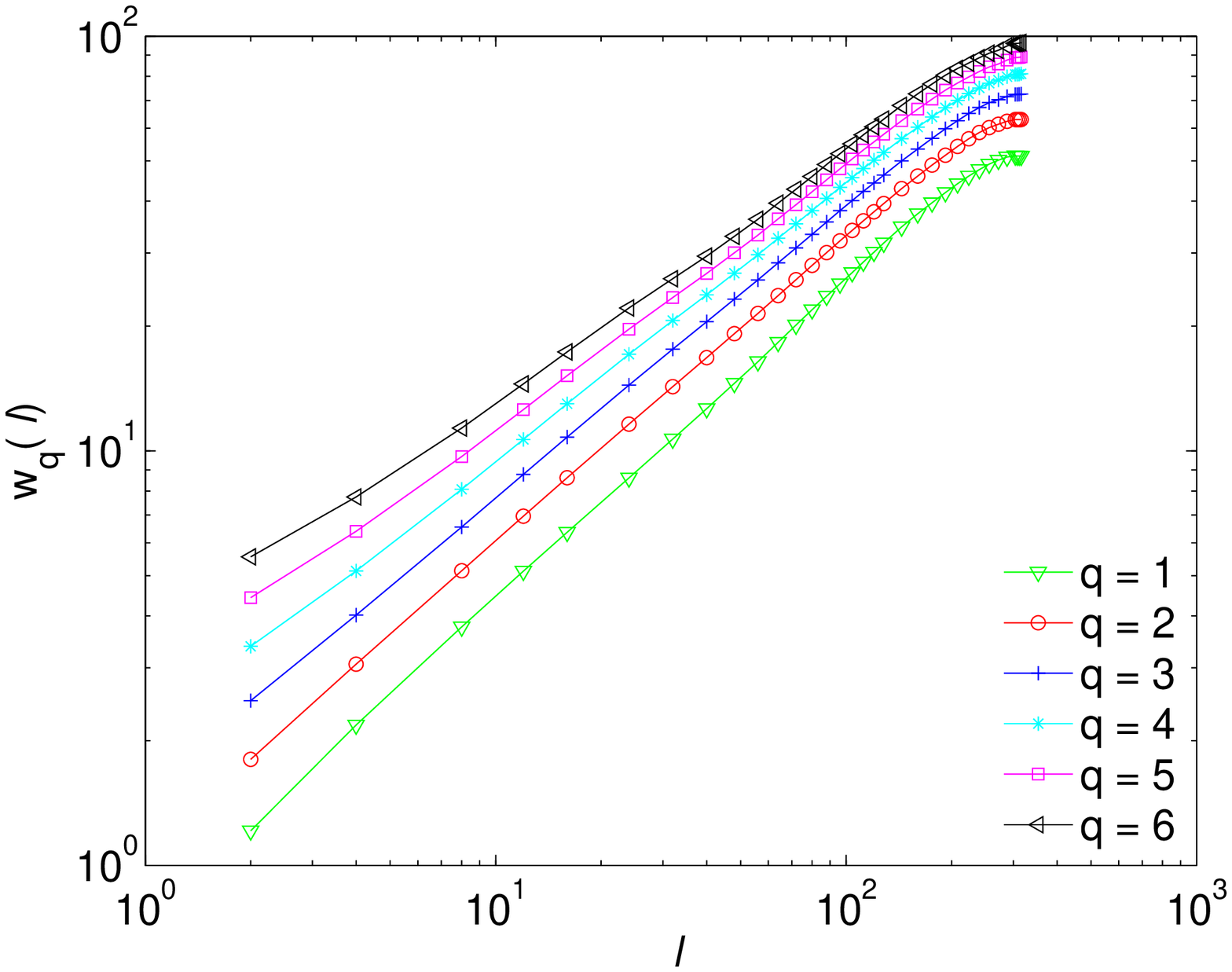}
\includegraphics[width=8cm]{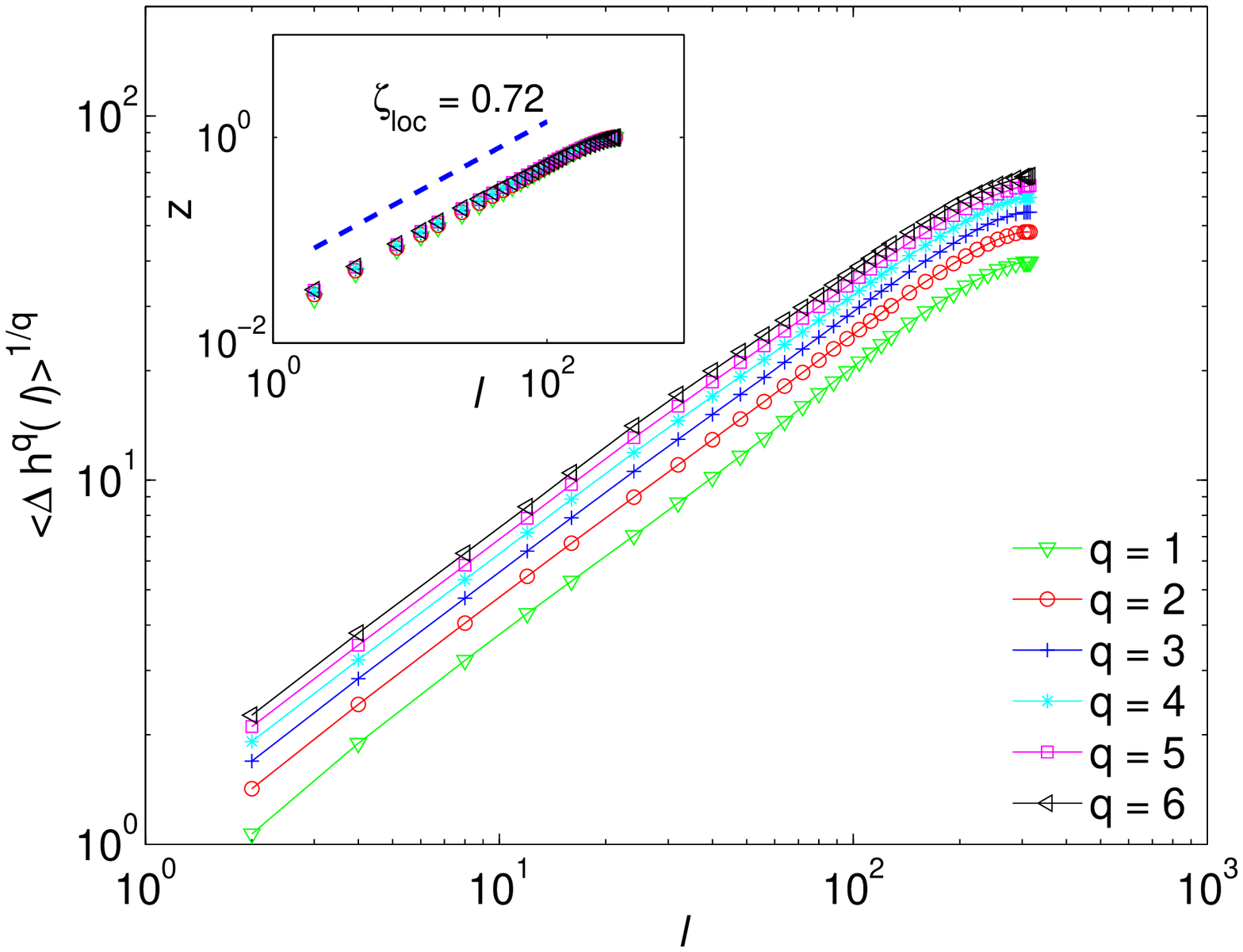}
\caption{(Color online) Scaling of $q$-th order
correlation function $C_q(\ell)$. The data presented is for a system of size $L = 320$ simulated
with open boundary conditions.
(a) $C_q(\ell)$ measured using original crak profiles with jumps.
(b) $C_q(\ell)$ measured using crack profiles without jumps.
Removal of overhangs in the crack profile eliminates
apparent multiscaling. Inset shows that normalization of the
data leads to collapse of the curves with a local roughness
exponent $\zeta_{loc} = 0.72$.}
\label{fig:nopbc_nomultiscaling}
\end{figure}

On the other hand, removing the jumps in the crack profiles once again
completely eliminates the multiscaling. Figure
\ref{fig:nopbc_nomultiscaling}a presents the scaling of $q$-th order
correlation function $C_q(\ell) = \langle
|h(x+\ell)-h(x)|^q\rangle^{1/q}$ measured using the original crack
profiles. Multiscaling below a characteristic length scale can be
clearly seen in Fig. \ref{fig:nopbc_nomultiscaling}a. The data in Fig.
\ref{fig:nopbc_nomultiscaling}b represents the scaling of $q$-th order
correlation function $C_q(\ell)$ measured after removing the jumps
in the profiles. The Figure shows that the plots for different crack
profile moments $q$ are parallel to one another, and thus the
removal of jumps in the crack profiles eliminates multiscaling. A
collapse of these plots is shown in the inset and the local
roughness exponent is estimated to be $\zeta_{loc} = 0.72$, close to
the PBC value.

%nopbc

% nopbc

\begin{figure}[hbtp]
\includegraphics[width=8cm]{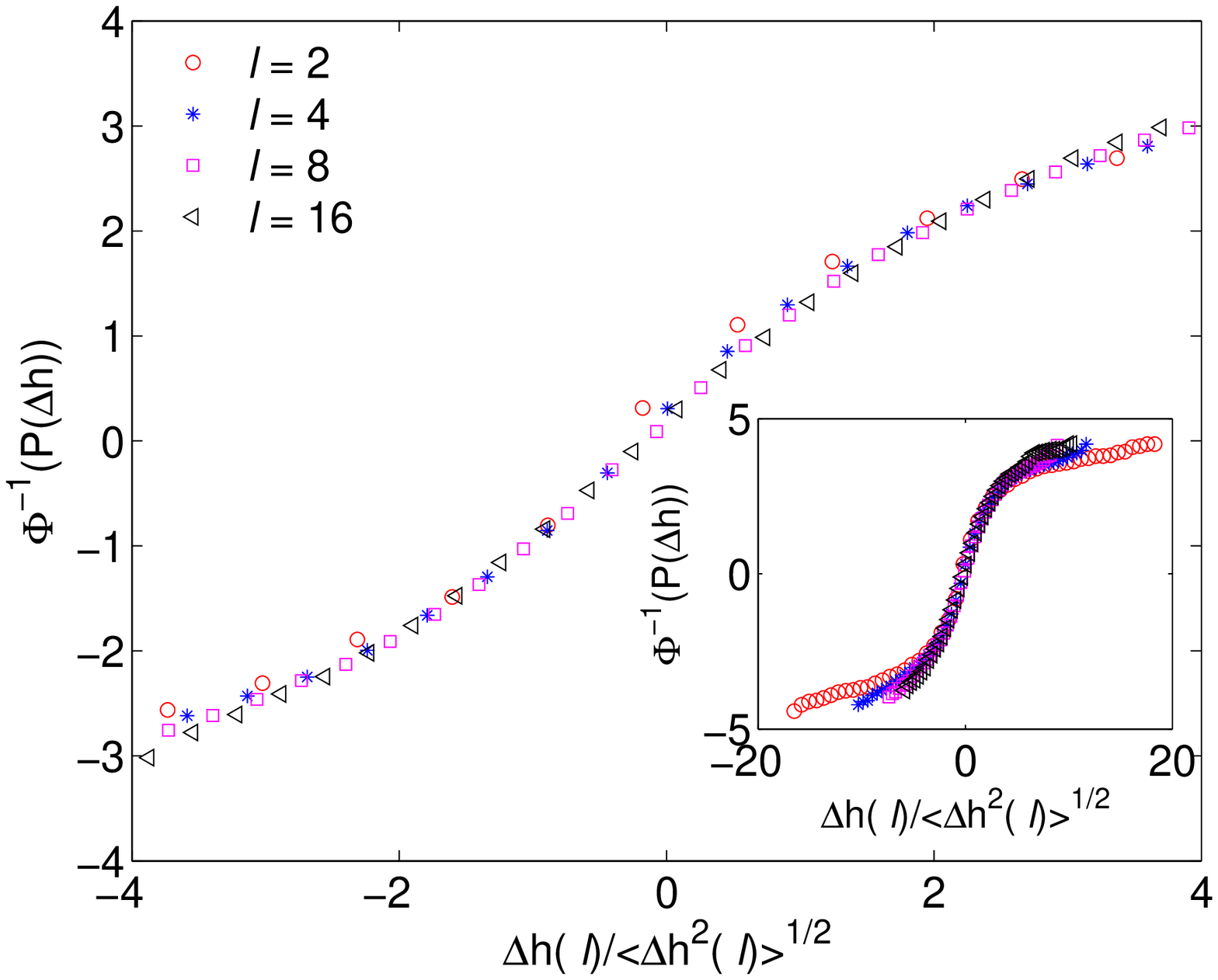}
\includegraphics[width=8cm]{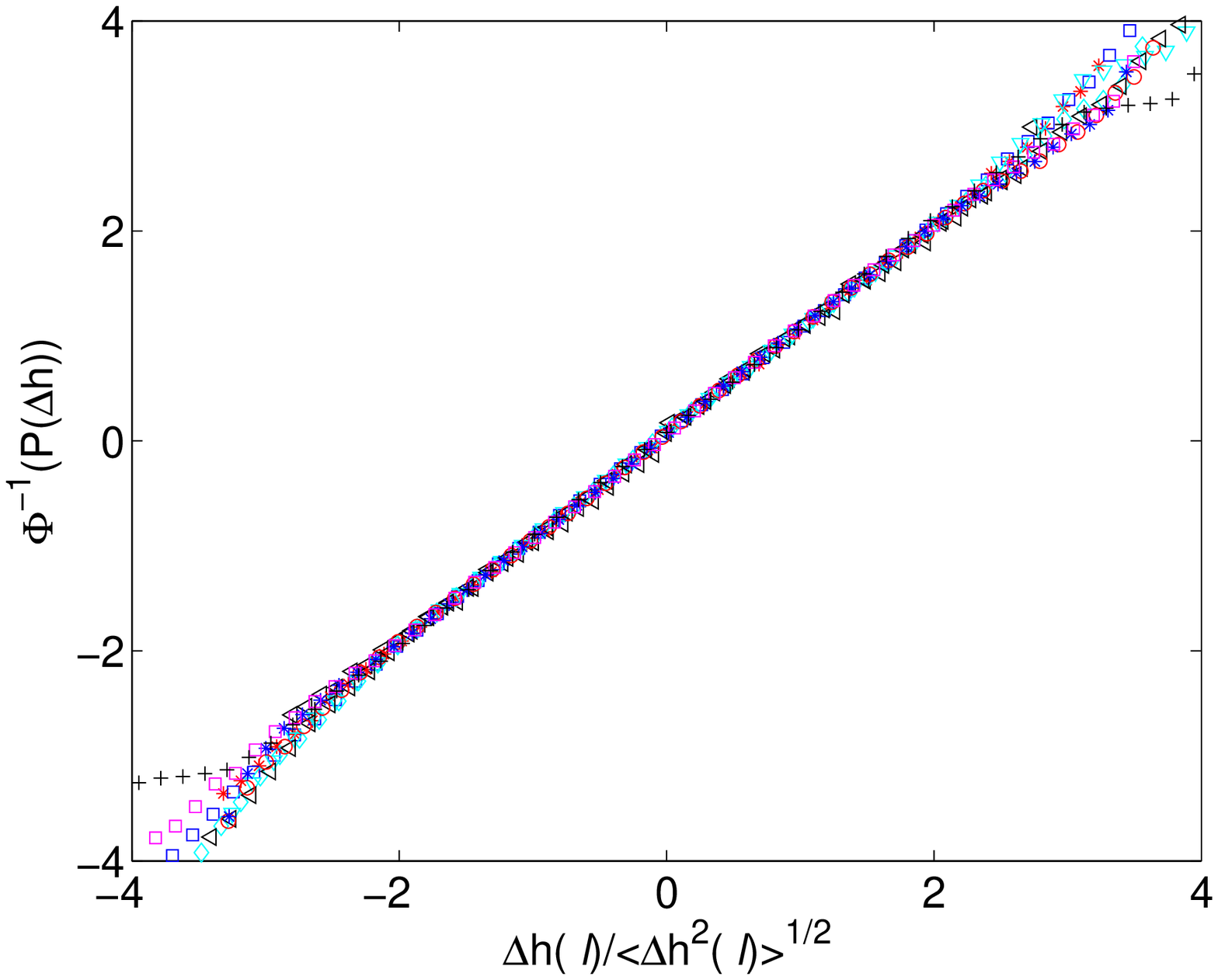}
\caption{(Color online) Plots of cumulative probability
distributions $P(\Delta h(\ell))$ of the height differences $\Delta
h(\ell) = [h(x+\ell) - h(x)]$ of the crack profile $h(x)$ for
various bin sizes $\ell$ on a normal paper. $\Phi^{-1}$ denotes
inverse Gaussian. The collapse of the profiles onto a straight line
with unit slope indicates that a Gaussian distribution is adequate
to represent $P(\Delta h(\ell))$. (a) $P(\Delta h(\ell))$
distributions for $L = 512$ and $\ell \ll L$. Large deviation from
Gaussian profiles is observed for these window sizes. (b) Removing
the jumps in the profiles however collapses the $P(\Delta h(\ell))$
distributions onto a straight line indicating the adequacy of a
Gaussian even for small window sizes $\ell$. The data is for $L =
512$ and $\ell = (2,4,8,16,32,64,96,128,160,256)$.}
\label{fig:pdeltah}
\end{figure}

In the following, we finally investigate the probability density
$p(\Delta h(\ell))$ of height differences $\Delta h(\ell)$. In Refs.
\cite{jstat2,santucci07}, the $p(\Delta h(\ell))$ distribution is
shown to follow a Gaussian distribution above a cutoff length scale
and the deviations away from Gaussian distribution in the tails of
the distribution have been attributed to finite jumps in the crack
profiles. A self-affine scaling of $p(\Delta h(\ell))$ as given by
Eq. (\ref{pdelt}) implies that the cumulative distribution $P(\Delta
h(\ell))$ scales as $P(\Delta h(\ell)) \sim P(\Delta
h(\ell)/\langle\Delta h^2(\ell)\rangle^{1/2})$. Figure
\ref{fig:pdeltah}(a) presents the raw data of cumulative probability
distributions $P(\Delta h(\ell))$ of the height differences $\Delta
h(\ell)$ on a normal or Gaussian paper for bin sizes $\ell \ll L$. As observed
in Refs. \cite{jstat2,santucci07}, and in Ref. \cite{salminen03}
Fig. \ref{fig:pdeltah}(a) shows large deviations away from Gaussian
distribution for these small bin sizes. However, for moderate bin
sizes, the distribution is Gaussian with deviations in the tails of
the distribution beyond the $3\sigma = 3\langle\Delta
h^2(\ell)\rangle^{1/2}$ limit (data not shown in Figure). Removing
the jumps in the crack profiles however collapses the $P(\Delta
h_{P}(\ell))$ distributions onto a straight line (see Fig.
\ref{fig:pdeltah}(b)) indicating the adequacy of Gaussian
distribution even for small window sizes $\ell$. Indeed, Fig.
\ref{fig:pdeltah}(b) shows the collapse of the $P(\Delta
h_{P}(\ell))$ data for a system size $L = 512$ with a variety of bin
sizes $2 \le \ell \le L/2$. Removing the jumps in the profiles not
only made the $P(\Delta h_{P}(\ell))$ distributions Gaussian even
for small window sizes $\ell$ but also extended the validity of
$P(\Delta h_{P}(\ell))$ Gaussian distribution for moderate bin sizes
to a $4\sigma = 4\langle\Delta h_{P}^2(\ell)\rangle^{1/2}$
($99.993\%$ confidence) limit.

% unnotched, nojumps

\section{Discussion}
In summary, we have here considered the nature of the roughness of
the crack surfaces in the two-dimensional RFM. The results presented here 
indicate universality of local roughness exponent for both notched and
unnotched samples with different disorders $D$ in the range $0.3 \le
D \le 1.0$ and for different relative crack sizes $a_0/L$. This is
true both for open and periodic boundary conditions.

The results indicate that anomalous scaling of roughness is a
generic feature of two-dimensional fracture in the fuse model. This
is in contrast to e.g. the beam model \cite{newbeam}, where the
global and local exponents are equal. The difference of the global and
local exponents arises due to an additional lengthscale, which scales as
a power-law of the system size $L$. We further investigated whether
anomalous scaling of roughness is an artifact of presence of large
jumps in the tails of $p(\Delta h(\ell))$ distribution. 
To do this, we considered the $\Delta h(\ell)$ data that is only within 
$\pm 3\sigma$ range of mean of $p(\Delta h(\ell))$ distribution, and 
computed the corresponding $\langle\Delta h^2(\ell)\rangle^{1/2}$ for 
various window sizes $\ell$. However, the data even from these truncated 
$p(\Delta h(\ell))$ distributions showed anomalous scaling. We repeated 
our investigation with $\pm 2\sigma$ range as well, but with a similar result. 
This indicates that anomalous scaling of roughness is not due to the 
tails of $p(\Delta h(\ell))$ distribution and persists in the mean 
as a function of $L$. 

Our results provide a concrete proof that the apparent
multi-scaling of crack profiles observed in Ref. \cite{procaccia} is
an artifact of jumps in the crack profiles that are formed due to
the solid-on-solid approximation used in extracting the crack
profiles. The removal of these jumps from the crack profiles
completely eliminates this apparent multi-scaling of crack profiles.
Furthermore, removing these jumps in the crack profiles extends the
validity of Gaussian probability density distribution $p(\Delta
h(\ell))$ of the height differences to even smaller window sizes
$\ell$ and to a range (of $4\sigma = 4\langle\Delta
h_{P}^2(\ell)\rangle^{1/2}$) well beyond that observed in earlier
studies.

In conclusion, though the  RFM is a "toy model" of (two-dimensional
here) fracture, it still poses interesting issues and can be used to
study questions that are also relevant for experiments. Our numerical 
results presented here raise three basic theoretical questions related to 
the morphology of two-dimensional RFM fracture surfaces that still remain 
to be answered. First, why does the extra lengthscale that leads to
anomalous scaling have to be algebraic? Second, models explaining
the dynamics in the final avalanche (unstable crack propagation) and
the roughness exponent values would be theoretically interesting to
develop. Third, how strong is the universality of the roughness exponents for
disorders very different from the ones used here, in particular, to those leading to
percolative damage or finite densities of infinitely strong fuses?
Finally, in addition to presenting results that explain the origins of apparent
multiscaling and anomalous scaling, we have also made a connection
between periodic fracture surfaces and excursions of
stochastic processes.

\par
\vskip 1.00em 
\noindent 
{\bf Acknowledgment} \\
This research is sponsored by the Mathematical, Information and Computational
Sciences Division, Office of Advanced Scientific Computing Research,
U.S. Department of Energy under contract number DE-AC05-00OR22725
with UT-Battelle, LLC. MJA and SZ gratefully thank the financial
support of the European Commissions NEST Pathfinder programme TRIGS
under contract NEST-2005-PATH-COM-043386. MJA also acknowledges the
financial support from The Center of Excellence program of the
Academy of Finland, and the hospitality of the Kavli Institute of
Theoretical Physics, China in Beijing.

\end{document}